\documentclass[12pt,a4paper]{article}
\usepackage{cite}
\usepackage{fullpage}
\usepackage[centertags]{amsmath}
\allowdisplaybreaks[4]
\usepackage{amssymb}
\usepackage{bm}
\usepackage[dvips]{graphicx}
\usepackage{url}
\usepackage[hyperindex]{hyperref}
\usepackage{color}

\newcommand{\half}{\frac{1}{2}}
\newcommand{\Vcc}{V_{\mathrm{CC}}}
\newcommand{\Vnc}{V_{\mathrm{NC}}}
\newcommand{\Rnc}{R_{\mathrm{NC}}}

\newcommand{\dM}{\delta_{\mathrm{M}}}
\newcommand{\dMk}{\delta_{\mathrm{M}(k)}}

\newcommand{\Ptot}[3]{P^{\mathrm{#1}}_{\nu_{#2} \to \nu_{#3}}}

\newcommand{\averP}[2][e]{\overline{P}^{\mathrm{#2}}_{\nu_{#1} \to \nu_\beta}}
\newcommand{\Ptb}[1][E]{P^{\mathrm{#1}}_{\nu_2 \to \nu_\beta}}
\newcommand{\Pte}[1][E]{P^{\mathrm{#1}}_{\nu_2 \to \nu_e}}

\newcommand{\averR}[1][\beta]{\overline{R}_{2#1}}
\newcommand{\averD}[1][\beta]{\overline{D}_{e#1}}

\newcommand{\psiV}{\psi^{\mathrm{V}}}
\newcommand{\PsiV}{\Psi^{\mathrm{V}}}
\newcommand{\psiM}{\psi^{\mathrm{M}}}
\newcommand{\PsiM}{\Psi^{\mathrm{M}}}

\newcommand{\HV}{\mathcal{H}^{\mathrm{V}}}
\newcommand{\HM}{\mathcal{H}^{\mathrm{M}}}

\newcommand{\HMad}{\HM_{\mathrm{ad}}}
\newcommand{\HMna}{\HM_{\mathrm{na}}}
\newcommand{\SV}{S^{\mathrm{V}}}
\newcommand{\SVX}{\SV(x_f,x_i)}
\newcommand{\SVslab}{\SV_{\mathrm{slab}}}
\newcommand{\SVslabk}{\SV_{\mathrm{slab}(k)}}

\newcommand{\SM}{S^{\mathrm{M}}}

\newcommand{\SMad}{\SM_\mathrm{ad}}
\newcommand{\SMX}{\SM(x_f,x_i)}
\newcommand{\SMXad}{\SM_\mathrm{ad}(x_f,x_i)}
\newcommand{\SMXada}{\SM_\mathrm{ad}(x,x_i)}

\newcommand{\W}[1][\omega]{W(#1,\varphi)}
\newcommand{\WD}[1][\omega]{W^\dagger(#1,\varphi)}

\newcommand{\T}{\mathrm{T}}

\newcommand{\dx}{\mathrm{d}x}
\newcommand{\ddx}{\frac{\mathrm{d}}{\mathrm{d} x}}
\newcommand{\df}{\dot{\varphi}}
\newcommand{\dw}{\dot{\omega}}
\newcommand{\intx}{\int_{x_i}^{x_f}}
\newcommand{\mIm}{\mathrm{Im}}
\newcommand{\mRe}{\mathrm{Re}}

\begin{document}

\vspace{1cm}
\begin{center}
{\large\bf Day-Night Asymmetries in Active-Sterile Solar Neutrino
Oscillations}
\end{center}
\vspace{0.3cm}

\begin{center}
{\bf H.W. Long} \footnote{E-mail: lhw0128@mail.ustc.edu.cn}\\
Department of Modern Physics, University of Science and\\
Technology of China, Hefei, Anhui 230026, China \\
\vspace{0.5cm}
{\bf Y.F. Li} \footnote{E-mail: liyufeng@ihep.ac.cn}\\
 Institute of High Energy Physics, Chinese Academy of\\
Sciences, Beijing 100049, China \\
\vspace{0.5cm}
{\bf C. Giunti}  \footnote{E-mail: giunti@to.infn.it}\\
INFN, Sezione di Torino, Via P. Giuria 1, I--10125 Torino, Italy\\
\end{center}

\setcounter{footnote}{0}

\vspace{1.5cm}
\begin{abstract}
Day-night asymmetries in active-sterile solar neutrino oscillations
are discussed in the general $3+N_{s}$ mixing framework with
three active and $N_s$ sterile neutrinos.
Analytical expressions of
the probability of neutrino flavor transitions
in the Earth
in the perturbative approximation
and in the slab approximation are presented
and
the effects of active-sterile mixing and of the CP-violating phases
are discussed.
The accuracy of the analytical approximations and the properties of
the day-night asymmetries are illustrated numerically
in the 3+1 neutrino mixing framework.

\end{abstract}
\vspace{1.5cm}
\newpage

\section{Introduction}
\label{introduction}

After the measurement of non-zero $\theta_{13}$ in recent reactor
\cite{DYB,RENO,DC} and accelerator \cite{T2K,MINOS} neutrino
experiments, we have a well-established standard framework of
three-neutrino oscillations, which explains solar, atmospheric,
reactor and accelerator neutrino oscillation data \cite{PDG} by two
distinct mass-squared differences (i.e.,
$\Delta{m}^{2}_{\mathrm{SOL}}$ and $\Delta{m}^{2}_{\mathrm{ATM}}$)
and three non-zero mixing angles ($\theta_{12}$, $\theta_{23}$ and
$\theta_{13}$). However, there are some short base-line (SBL)
neutrino oscillation anomalies that cannot be explained in the
three-neutrino mixing framework: the anomalies found in the
LSND \cite{LSND} and MiniBooNE \cite{Mini}
accelerator experiments, the reactor antineutrino anomaly \cite{R} and the
Gallium anomaly \cite{Ga1,Ga2}.
A neutrino oscillation explanation of these anomalies implies
the existence of
at least one extra mass-squared difference $\Delta{m}^{2}_{\mathrm{SBL}}$,
such that $\Delta{m}^{2}_{\mathrm{SOL}}
\ll \Delta{m}^{2}_{\mathrm{ATM}} \ll
\Delta{m}^{2}_{\mathrm{SBL}}$, and a small mixing of the three active
neutrinos with extra sterile neutrino states
\cite{Ga2,Schwetz4,Giunti11,whitepaper}. Moreover, analyses
\cite{Raffelt,Giunti12,Mangano} of the data from cosmological
observations and Big-Bang Nucleosynthesis may point to the existence
of additional dark radiation in the Universe, for which light
sterile neutrinos are one of the most promising candidates.

The existence of light sterile neutrinos and their mixing with the
standard active neutrinos can be tested also by studying their
effects in solar \cite{solarste,CP,palazzo1,palazzo2,smirnov} and
atmospheric \cite{iceatm} neutrino oscillations, as well as in the
processes of the beta decay \cite{beta_exp,beta_ph} and
neutrino-less double beta decay \cite{double}. In this paper we
discuss the possible effects of active-sterile mixing on the
day-night asymmetries of solar neutrino oscillations.

Solar neutrinos are produced in the core of the Sun and undergo the
large-mixing-angle (LMA) Mikheev-Smirnov-Wolfenstein (MSW) mechanism
\cite{Wolfenstein:1978ue,Mikheev:1985gs} before they propagate out
of the solar surface. This mechanism was established about ten years
ago after the decisive evidences in favor of solar neutrino
oscillations obtained in the SNO \cite{SNOnc} and KamLAND
\cite{KL02} experiments. Since then, testing the MSW mechanism with
low energy solar neutrinos \cite{BX} and searching for sub-leading
effects \cite{solarste,CP,palazzo1,palazzo2} beyond the standard
three-neutrino oscillation have become the central concern of solar
neutrino searches. Moreover, the terrestrial matter effect in terms
of the regeneration effects and day-night asymmetries \cite{DN} of
solar neutrino oscillations can give direct and independent tests of
the MSW mechanism and constrain different kinds of sub-leading
effects. Different calculations of the day-night asymmetries in the
three-neutrino mixing scheme have been discussed in the recent
literature \cite{blennow,Valle,liao,Aleshin}. In this work we extend
the discussion to the general framework with three active and $N_s$
sterile neutrinos (the $3+N_s$ scheme). In this study, we derive the
analytical expressions of the day-night asymmetries in the $3+N_s$
mixing scheme without any constraint on the mixing elements,
assuming only a realistic hierarchy on the mass-squared differences.
Calculations using both the perturbative approximation and the slab
approximation are checked by comparing their results with those of
accurate numerical calculations.

The outline of this work is planned as follows. In Section~\ref{sec:review} we
review briefly the general framework of neutrino evolution in matter
and present the neutrino oscillation probabilities relevant to the
experimental results in the daytime \cite{CP}. In Section~\ref{sec:P_SE} we
calculate the oscillation probabilities in the nighttime for
solar neutrinos propagating inside the Earth before being detected,
and we discuss the calculation of the day-night asymmetries by using different
approximation methods. In Section~\ref{sec:Numerical} we test the accuracy of our
analytical calculations in different approximations
comparing the results with those obtained with an accurate numerical
solution of the solar neutrino evolution inside the Earth.
We also illustrate
the properties of the day-night asymmetries using the analytical approximations.
Finally, we
conclude in Section~\ref{sec:Conclusion} and we present in the Appendices additional
details of the calculations
and the
parametrization of the four-neutrino mixing matrix that we use in the examples.

\section{Oscillations from the Sun to Earth}
\label{sec:review}

In this Section, we review the general framework in Ref.~\cite{CP}
of the neutrino evolution in matter and the solar neutrino
oscillation probabilities in the daytime.

To begin with, the evolution of solar neutrinos propagating in
matter (in the Sun or in the Earth) is described by the MSW
equation\cite{Wolfenstein:1978ue,Mikheev:1985gs}
\begin{equation}
  i \ddx \Psi = \left( U \mathcal{M}
    U^{\dagger} + \mathcal{V} \right) \Psi
  \quad, \label{eq:MSW_eq}
\end{equation}
where $\Psi=\left(
  \psi_{e},\psi_{\mu},\psi_{\tau},\psi_{s_{1}},\ldots,\psi_{s_{N_{s}}}
\right)^T$ is the flavor transition amplitudes, $U$ is the $(3+
N^{}_s)\times (3+ N^{}_s)$ neutrino mixing matrix, and
\begin{align}
  \mathcal{M} = \null & \null \mathrm{diag} \! \left( 0,
    \frac{\Delta{m}^2_{21}}{2 E}, \frac{\Delta{m}^2_{31}}{2 E}, \frac{\Delta{m}^2_{41}}{2 E},
    \ldots,\frac{\Delta{m}^2_{N1}}{2 E}\right)
  \quad, \\
  \mathcal{V} = \null & \null \mathrm{diag} \! \left(
    \Vcc+\Vnc, \Vnc, \Vnc,
    0, \ldots ,0 \right)
  \quad,\label{eq:potent}
\end{align}
where $E$ is the neutrino energy and $ \Delta{m}^2_{kj} = m_{k}^2
- m_{j}^2\,$ are the mass-squared differences. The charge-current (CC)
and neutral-current (NC) neutrino matter potentials in Eq.~(\ref{eq:potent}) are defined as
\begin{equation}
  \Vcc=\sqrt{2}G_{\mathrm{F}}N_{e}\simeq 7.63 \times 10^{-14}
  \, \frac{ N_{e} }{ N_{\mathrm{A}} \, \mathrm{cm}^{-3} } \,\, \mathrm{eV}
  \quad, \qquad
  \Vnc=-\half\sqrt{2}G_{\mathrm{F}}N_{n}
  \quad,
\end{equation}
where $G_{\mathrm{F}}$ is the Fermi constant, $N_{e}$ and $N_{n}$
are the electron and neutron number densities respectively, and
$N_{\mathrm{A}}$ is the Avogadro's number. In a neutral medium, we
can use the electron fraction
\begin{equation}
Y_{e} = \frac{N_{e}}{ N_{e}+N_{n}}\quad,
\end{equation}
to define the ratio $\Rnc$ of matter potentials as
\begin{equation}
  \Vnc = \Rnc \Vcc \quad, \quad \mathrm{with}
  \quad \Rnc = - \frac{ 1 - Y_{e} }{ 2 Y_{e} }
  \quad. \label{eq:Rnc}
\end{equation}

In the vacuum mass basis $ \PsiV = \left(\psiV_{1},\ldots, \psiV_{N}
\right)^T = U^{\dagger} \Psi\,$, the flavor
transitions generated by $\Delta{m}^2_{21} $ are decoupled from those generated by
the larger mass-squared differences, because of the hierarchy
\begin{equation}
  V_{\mathrm{CC}} \sim |V_{\mathrm{NC}}| \lesssim  \frac{\Delta{m}^2_{21}}{2 E}
  \ll \frac{ |\Delta{m}^2_{k1}| }{2 E} \quad \mathrm{for} \quad k \geq 3
  \label{eq:Hierarchy}
  \quad,
\end{equation}
for solar and terrestrial matter densities. Therefore, the full
neutrino evolution equation in Eq.~(\ref{eq:MSW_eq}) can be
truncated to a $2\times2$ evolution equation
\begin{align}
  i \ddx \PsiV_2 &= \HV_2 \PsiV_2
  \quad, \label{eq:evol_eqV}
\end{align}
with $\PsiV_2=(\psiV_1,\psiV_2)^\T$ and
\begin{align}
  \HV_2=
  \begin{pmatrix}
    - \delta + V \cos 2 \xi & V \sin 2 \xi e^{i\varphi} \\
    V \sin 2 \xi e^{-i\varphi} & \delta - V \cos 2 \xi
  \end{pmatrix}\quad.
 \label{eq:HV}
  \end{align}
All the other amplitudes evolve independently as
\begin{align}
  \psiV_{k}(x) &\simeq \psiV_{k}(0) \, \exp\!\left(
    - i \, \frac{ \Delta{m}^2_{k1} x }{ 2 E } \right) \quad, \quad
  \mathrm{for} \quad k \geq 3
  \quad. \label{eq:3more}
\end{align}
The vacuum oscillation wave number $\delta$, the effective potential
$V$, the matter-dependent mixing angle $\xi$ and the complex phase $\varphi$
are defined respectively as
\begin{align}
  \delta &= \frac{\Delta m^2_{12}}{4E}
  \quad, \\
  V &= \half \Vcc \sqrt{ X^2 + |Y|^2 }
  \quad, \\
  \xi & = \half \arctan \frac{|Y|}{X}
  \quad, \label{eq:xi} \\
  \varphi &= \arg(Y)
  \quad, \label{eq:varphi}
\end{align}
with
\begin{align}
  X &= |U_{e1}|^2 - |U_{e2}|^2 - \Rnc \sum_{i=1}^{N_{s}}\left(
    |U_{s_{i}1}|^2 - |U_{s_{i}2}|^2 \right)\label{eq:X}
  \quad, \\
  Y &= 2 \left( U_{e1}^{*} U_{e2} - \Rnc \sum_{i=1}^{N_{s}}
    U_{s_{i}1}^{*} U_{s_{i}2} \right)
  \quad. \label{eq:Y}
\end{align}
We can define an effective mass basis in matter for the
two-neutrino framework in Eq.~(\ref{eq:evol_eqV}) as
\begin{align}
  \PsiV_2 &= W_2(\omega,\varphi) \PsiM_2
  \quad, \label{eq:psiMtoV}
\end{align}
where $\PsiM_2=(\psiM_1,\psiM_2)^\T$ is the amplitude vector in the
effective mass basis, and
\begin{align}
  W_2(\omega,\varphi) &\equiv
  \begin{pmatrix}
    \cos\omega & \sin\omega e^{i \varphi} \\
    -\sin\omega e^{-i \varphi} & \cos\omega
  \end{pmatrix}
  \quad,\label{eq:W}
\end{align}
with
\begin{align}
  \tan 2\omega &= \frac{ V \sin2\xi }{ \delta - V \cos2\xi }
  \quad, \label{eq:omega}
\end{align}
is the unitary matrix which diagonalizes the Hamiltonian in Eq.~(\ref{eq:HV}).
The evolution equation in the effective mass basis in matter is
\begin{align}
  i \ddx \PsiM_2 &= \HM_2 \PsiM_2
  \quad, \label{eq:evol_eqM}
\end{align}
where the Hamiltonian is given by
\begin{align}
  \HM_2 &= \HMad + \HMna
  \nonumber{} \\
  & \equiv
  \begin{pmatrix}
    -\dM & 0
    \\
    0 & \dM
  \end{pmatrix}
  +
  \begin{pmatrix}
    - \df \sin^2\omega & ( \half \df \sin2\omega - i \dw ) e^{i \varphi}
    \\
    ( \half \df \sin2\omega + i \dw ) e^{-i \varphi} & \df \sin^2\omega
  \end{pmatrix}
  \quad, \label{eq:HM}
\end{align}
with
\begin{align}
  \dM &= \sqrt{ (\delta - V \cos2\xi)^2 + (V \sin2\xi)^2 }
  \quad, \label{eq:dM}\\
  \df &\equiv \frac{\mathrm{d} \varphi}{\mathrm{d} x} \quad , \quad
  \dw \equiv \frac{\mathrm{d} \omega}{\mathrm{d} x}
  \quad.
\end{align}
Note that we have decomposed the Hamiltonian into the adiabatic (ad)
and non-adiabatic (na) parts in Eq.~(\ref{eq:HM}),
which take into account the dependence of $\omega$ and $\varphi$
on the variable matter density along the neutrino propagation path.

To proceed, we define a parametrization \cite{solarste} of the
neutrino mixing matrix $U$ as
\begin{equation}
  \left\{\begin{array}{ll}
      U_{\beta 1}&=\cos\theta_\beta \, \cos\chi_\beta \, e^{i\,\phi_{\beta 1}}\\
      U_{\beta 2}&=\sin\theta_\beta \, \cos\chi_\beta \, e^{i\,\phi_{\beta 2}}
    \end{array}\right.\quad
  \text{with} \quad \cos^2\chi_{\beta} = |U_{\beta 1}|^2 + |U_{\beta 2}|^2
  \quad, \label{eq:U_theta_chi}
\end{equation}
and a formal solution of the averaged amplitudes for solar neutrino
evolution as
\begin{align}
  \left\{
  \begin{array}{ll}
    \overline
    { |\psiM_1(x_d)|^2 } &= |\psiM_1(0)|^2 \left( 1 - P_{12} \right) + |\psiM_2(0)|^2 P_{12}
    \\
    \overline
    { |\psiM_2(x_d)|^2 } &= |\psiM_1(0)|^2 P_{12} + |\psiM_2(0)|^2 \left( 1 - P_{12} \right)
  \end{array} \right.  \label{eq:P12def}
\quad\,,
\end{align}
where $x_d$ is the coordinate of the detector on the Earth, $P_{12}$
is the level-crossing probability generated by the off-diagonal
non-adiabatic terms in Eq.~(\ref{eq:HM}) between $\psiM_1$ and
$\psiM_2$ during their propagation inside the Sun.

Finally, we obtain the averaged solar neutrino oscillation
probabilities in the daytime (see Ref.~\cite{CP} for a careful
derivation) as
\begin{align}
  \averP{S} &= \cos^2\chi_e \cos^2\chi_{\beta} \averP{(2\nu)} + \sum_{k=3}^{N} |U_{ek}|^2 |U_{\beta k}|^2
  \quad, \label{eq:averP_S}
\end{align}
where
\begin{align}
  \averP{(2\nu)} &=  \half + (\half - P_{12}) \cos2\Theta_{e}^{0} \cos2\theta_{\beta}
  \quad, \label{eq:averPS2f} \\
  \cos2\Theta_{e}^{0} &= \cos2\theta_e \cos2\omega^0 -
  \cos\Phi^0_e \sin2\theta_e \sin2\omega^0\quad,\\
  \Phi^0_e&=\phi_{e1} - \phi_{e2} +\varphi^0
  \quad,
\end{align}
and all the parameters with a `0' superscript are defined at the
neutrino production point.

\section{Day-Night Asymmetries}
\label{sec:P_SE}

In Section~\ref{sec:review} we have briefly reviewed the derivation
of the solar neutrino oscillation probabilities (\ref{eq:averP_S})
in the daytime. However, solar neutrino fluxes might be affected by
the Earth's matter if they are detected at night and travel inside
the Earth before detection (the so-called ``regeneration effect'').
Therefore, there can be measurable differences of the oscillation
probabilities in the daytime and nighttime which can be quantified
by defining suitable day-night asymmetries. In the standard
framework of three-neutrino mixing, there is only a single day-night
asymmetry for the electron neutrino survival probability
\cite{blennow,Valle,liao,Aleshin}. In the general $3+N_{s}$ mixing
scheme with three active and $N_{s}$ sterile neutrinos which we
consider there is an additional asymmetry for the
electron-to-sterile neutrino transition probability, which can be
revealed by precise neutral-current measurements.

Because of the spread of the solar neutrino production areas and
finite detector energy resolution, the contributions of the
different mass eigenstates to the oscillation probabilities are
incoherent \cite{washout}. Therefore, the oscillation probabilities
without [crossing only the Sun (S)] and with [crossing both the Sun
(S) and the Earth (E)] the terrestrial matter effects are given,
respectively, by\footnote{ Some aspects of the terrestrial matter
effects in the case of four-neutrino mixing have been already
discussed in Refs.~\cite{Dooling:1999sg,Giunti:2000wt,palazzo1}.}
\begin{align}
  \averP{S} &= \sum_{k=1}^{3+N_{s}} \Ptot{S}{e}{k}
  \Ptot{V}{k}{\beta}\quad,
  \\
  \averP{SE} &= \sum_{k=1}^{3+N_{s}} \Ptot{S}{e}{k} \Ptot{E}{k}{\beta}
  \quad,
\end{align}
where
\begin{align}
  \Ptb[V] = |U_{\beta 2}|^2 = \sin^2\theta_\beta \cos^2\chi_\beta
  \quad,
\end{align}
is the transition probability between $\nu_{k}$ and $\nu_{\beta}$ in
vacuum. From Eq.~(\ref{eq:3more}), we obtain
\begin{equation}
  \Ptot{S}{e}{k} \simeq | U_{ek} |^2\,,\quad
  \Ptot{E}{k}{\beta} \simeq | U_{\beta k}
  |^2\,,\quad (k \geq 3)\,,
\end{equation}
and due to the unitarity of oscillation probabilities, we have the
following relations
\begin{align}
  \Ptot{S}{e}{1} &= 1 - \sum_{k=3}^{3+N_{s}} |U_{ek}|^2 - \Ptot{S}{e}{2}
  \quad, \\
  \Ptot{E}{1}{\beta} &= 1 - \sum_{k=3}^{3+N_{s}} |U_{\beta k}|^2 - \Ptb
  \quad.
\end{align}
By using the above relations and the probabilities in Eq.
(\ref{eq:averP_S}), we obtain the neutrino oscillation probabilities
including the terrestrial matter effects as follows:
\begin{align}
  &\averP{SE} = \averP{S} + \left( \Ptb - \Ptb[V] \right)
  \nonumber \\
  &\times \Big\{\frac
  { (|U_{e1}|^{2}+|U_{e2}|^{2}) (|U_{\beta 1}|^{2}+|U_{\beta 2}|^{2})
    -2(\averP{S} - \sum_{k=3}^{N}|U_{ek}|^{2}|U_{\beta k}|^{2}) }
  { |U_{\beta 1}|^{2} - |U_{\beta2}|^{2} } \Big\}
  .\label{eq:averP_SE2}
\end{align}
Using Eqs.~(\ref{eq:P12def}), (\ref{eq:averP_S}) and (\ref{eq:averPS2f}),
this expression can be written as\footnote{From the discussion in Ref.~\cite{CP},
we know that $P_{12}$ is negligibly small in the Sun.
Therefore, we neglect it in the numerical discussion in Section~\ref{sec:Numerical}.}
\begin{align}
  \averP{SE} = \averP{S} - \cos^2\chi_e (1-2P_{12})
  \cos2\Theta^0_{e} R_{2\beta}
  \quad, \label{eq:averP_SE}
\end{align}
where
\begin{align}
R_{2\beta}= \Ptb - \Ptb[V]
  \quad, \label{eq:R2beta}
\end{align}
are the regeneration factors of active-sterile solar neutrino
oscillations.

It is convenient to define the day-night
asymmetries of the oscillation probabilities
\begin{align}
  D_{e\beta} = \averP{SE} - \averP{S}
  =- \cos^2\chi_e (1-2P_{12}) \cos2\Theta^0_{e} \, R_{2\beta}
  \quad, \label{eq:Debeta}
\end{align}
which depend on the mixing parameter $\chi_e$,
on the solar oscillations quantities $P_{12}$ and $\Theta^0_{e}$,
and on the regeneration factors
$R_{2\beta}$ in the Earth.

Note that,
taking into account the unitarity of probabilities,
the measurable event rates in solar experiments
depend on two
independent regeneration factors: $R_{2e}$ and $R_{2s}$.
To count the
number of relevant mixing angles and CP-violating phases in the
regeneration factors, we can employ the same arguments presented in
Ref.~\cite{CP}. Because of the freedom in the complex rotations of
the $\nu_\mu$-$\nu_\tau$ sector and among the sterile neutrino
flavors, the probability $\Ptb$ depends on
3$N^{}_s$+2 mixing angles and
2$N^{}_s$ Dirac CP-violating phases.
In order to express the regeneration factors and the oscillation probabilities in terms of
this minimal number of oscillation parameters,
it is necessary to choose a convenient parameterization of the mixing matrix.
However, considering for example the 3+1 mixing scheme,
it is interesting to study the information that
solar neutrino measurements
can give on the elements
$U_{e4}$, $U_{\mu4}$ and $U_{\tau4}$
of the mixing matrix,
which quantify the mixing of active and sterile neutrinos.
Therefore,
it is convenient to use the parametrization of $U$ in Eq.~(\ref{eq:Uparam}),
in which $U_{e4}$, $U_{\mu4}$ and $U_{\tau4}$
are proportional, respectively, to the sines of the mixing angles
$\theta_{14}$, $\theta_{24}$ and $\theta_{34}$
and
$U_{\mu4}$ and $U_{\tau4}$
are directly related, respectively,
to the CP-violating phases
$\eta_{24}$ and $\eta_{34}$
[see Eqs.~(\ref{Ue34}) and (\ref{Um4})].
In this case,
the regeneration factors
and the oscillation probabilities depend on all the six mixing angles and the three
CP-violating phases in Appendix~\ref{sec:paramU}\footnote{
The mixing angles and CP-violating phases
of the appropriate parameterization in which the
regeneration factors
and the oscillation probabilities
depend on only
five mixing angles and two CP-violating phases
are complicated functions of the mixing parameters in Appendix~\ref{sec:paramU}.
}.

The regeneration factors are determined by the transition
probabilities $\Ptb$ inside the Earth. By using Eqs.~(\ref{eq:3more})
and (\ref{eq:U_theta_chi}) and the initial condition $\psiV_k(x_i) =
\delta_{k2}$, we obtain the formal expression
\begin{align}
  \Ptb(x_d) =& | \psi_\beta(x_d) |^2 = | U_{\beta 1} \psiV_1(x_d) + U_{\beta 2} \psiV_2(x_d) |^2
  \nonumber{}\\
  =& \half \cos^2\chi_\beta \Big\{ 1 + \cos2\theta_\beta ( |\psiV_1(x_d)|^2 - |\psiV_2(x_d)|^2 )
  \nonumber{}\\
  & + 2 \sin2\theta_\beta \cos(\phi_{\beta 2} - \phi_{\beta 1}) \mRe \left( {\psiV_1}^*(x_d) \psiV_2(x_d) \right)
  \nonumber{}\\
  &- 2 \sin2\theta_\beta \sin(\phi_{\beta 2} - \phi_{\beta 1}) \mIm \left( {\psiV_1}^*(x_d) \psiV_2(x_d) \right) \Big\}
  \quad, \label{eq:formal_ptb}
\end{align}
where the amplitudes in the vacuum mass basis $\psiV_i(x_d)$ with
$(i=1,2)$ can be calculated using the evolution equation inside the
Earth.

The most accurate calculation of the oscillation probabilities
can be obtained with a numerical solution of the evolution equation
(\ref{eq:evol_eqV}), using a precise density profile of matter in the Earth.
However,
such a numerical solution is too time-consuming
if one wants to explore a large volume of the space of the mixing-parameters
and
often one needs analytic expressions for the oscillation probabilities in order to study
their properties.
Therefore,
in the following two subsections, we describe two different approximations
which allow us to
obtain analytical expressions of $\Ptb$.
We discuss the accuracy of these approximations in Section~\ref{sec:Numerical}.

\subsection{Perturbative Approximation}
\label{sec:P_E_2beta}

For the density profile inside the Earth, the matrix elements of the
non-adiabatic Hamiltonian in Eq.~(\ref{eq:HM}) are much smaller than
those of the adiabatic Hamiltonian and can be treated as a perturbation.
This is due to the fact that the effective potential $V$ itself can be treated
as a perturbation, because $V\ll{\Delta{m}^2_{21}}/{2 E}$.
Therefore, both $\omega$ and $\dot{\omega}$ are small,
even at the boundaries of two adjacent shells in the Earth
where there are sudden changes of the matter density.

In the $S$-matrix formalism, the neutrino evolution can be written
formally as
\begin{align}
  \Psi^{\mathrm{B}}_2(x_f) &= S^{\mathrm{B}}(x_f,x_i) \Psi^{\mathrm{B}}_2(x_i)
  \quad, \label{eq:S_PsiVorM}
\end{align}
where the superscript
denotes the mass basis in vacuum
($\mathrm{B}=\mathrm{V}$)
or in matter ($\mathrm{B}=\mathrm{M}$),
and $x_i$ and $x_f$ are the initial and final points of
the neutrino trajectory. The two $S$-matrices in the vacuum and matter mass bases
are connected by the transformation
\begin{align}
  \SVX = W_2(\omega_f,\varphi_f)\SMX
  W_2^{\dagger}(\omega_i,\varphi_i)\quad,\label{eq:SMtoV}
\end{align}
with $W_2(\omega,\varphi)$ given in Eq.~(\ref{eq:W}).

We can calculate the $S$-matrix in the mass basis in matter using
the general perturbation theory (see
Refs.~{\cite{Akhmedov-2004,Ioannisian}} for details):
\begin{align}
  \SMX & \simeq \SMXad - i \SMXad \intx \SMXada^{-1} \HMna(x) \SMXada \dx
  \nonumber{}\\
  &=\SMXad -i \SMXad
  \begin{pmatrix}
    -A & C \\
    C^{*} & A
  \end{pmatrix}
  \nonumber{} \\
  &=
  \begin{pmatrix}
    (1+iA) e^{i\Delta} & -iC e^{i\Delta} \\
    -iC^* e^{-i\Delta} & (1-iA) e^{-i\Delta}
  \end{pmatrix}
  \quad,\label{eq:SMpX}
\end{align}
with
\begin{align}
  \SMXad &= {\rm exp}\large\{\, -i \intx \HMad(x) \dx \,\large\}
  \nonumber{} \\
  &=
  \begin{pmatrix}
    e^{i \Delta(x_f,x_i)} & 0 \\
    0 & e^{-i \Delta(x_f,x_i)}
  \end{pmatrix}
  \quad,\\
  \Delta(x_f,x_i) &= \intx \dM \dx \quad, \label{eq:Delta} \\
  A(x_f,x_i) &= \intx \df \sin^2 \omega \dx \quad, \label{eq:A} \\
  C(x_f,x_i) &= \intx ( \half \df \sin2\omega - i \dw ) e^{i [ \varphi -2 \Delta(x,x_i) ]} \dx
  \quad. \label{eq:C} 
\end{align}
Since the matter effects at the boundaries (i.e., $x_i$ and $x_f$)
of the neutrino path inside the Earth are negligible, we have $
W(\omega_i,\varphi_i) = W(\omega_f,\varphi_f) = 1 $ and $\SVX = \SMX
$. Therefore, as shown in Appendix~\ref{sec:detailed_pert},
the transition probability $\Ptb$ can be approximated
to the first order of $A$ and $C$ as
\begin{align}
  \Ptb(x_f) &\simeq \Ptb[V] + \cos^2\chi_\beta \sin2\theta_\beta
  \mIm \left[ C e^{ i (2\Delta + \phi_{\beta1} - \phi_{\beta2}) }
  \right]
  \nonumber{}\\
  &= \Ptb[V] + \cos^2\chi_\beta \sin2\theta_\beta \mIm \left\{\intx
  ( \half \df \sin2\omega - i \dw ) e^{i [ 2 \Delta(x_f,x) + \Phi_{\beta} ]}
  \dx\right\}
  \,,\label{eq:Ptb_pert_C}
\end{align}
where $\Phi_{\beta}=\phi_{\beta1} - \phi_{\beta2} + \varphi$. Notice
that $\Phi_{\beta}$ is invariant under the rephasing transformation
of $U_{\alpha k} \to e^{i \varphi_\alpha} U_{\alpha k} e^{i
\varphi_k}$, and gives the intrinsic phase-dependence of the
day-night asymmetries.
Furthermore,
as shown in Appendix~\ref{sec:detailed_pert},
in the approximation of a constant
electron fraction $Y_e$ \cite{Lisi1997} (i.e., $\df=0$), the
transition probability $\Ptb$ can be further simplified to the first
order of $V$ as
\begin{align}
  \Ptb(x_f) &\simeq \Ptb[V] + \cos^2\chi_\beta \sin2\theta_\beta \sin2\xi
  \intx V(x) \sin \left[ 2 \Delta(x_f,x) + \Phi_{\beta} \right] \dx
  \,. \label{eq:Ptb_pert_V}
\end{align}
In Appendix~\ref{sec:3-flavor} we derive the limit of this
expression for $\Pte(x_f)$ in the case of vanishing active-sterile
mixing in order to show that it agrees with that presented in the
case of standard three-neutrino mixing in Ref.~\cite{Valle}.

In Section~\ref{sec:Numerical} we will show that the perturbative
approximation gives an accurate description of
neutrino evolution inside the Earth.
However, since the integrations in
Eqs.~(\ref{eq:Ptb_pert_C}) and (\ref{eq:Ptb_pert_V}) are very
time-consuming, it is useful to investigate if there are other
approximate methods which give accurate and more rapid
solutions of the evolution equation.
In the next subsection we discuss the analytical
calculation of $\Ptb$ in the slab approximation.

\subsection{Slab Approximation}
\label{sec:Slab}

In the slab approximation,
the radial symmetric profile of the matter density inside the Earth is divided into $N$
shells ($1\leqslant i\leqslant N$, where $i=1$ is the innermost
shell), such that the density variation is very small within each shell
and sudden changes happen at the boundaries (the biggest change happens at the
mantle-core boundary).

When a solar neutrino travels inside the Earth, the neutrino
trajectory crosses the outer $n$ shells ($n\leqslant N$) of the
Earth and contains $2n-1$ segments with constant density. The number
$n$ and the lengths of the segments depend on the nadir angle of the
trajectory, which ends at the detector. We denote the coordinates of
the segment boundaries as $x_{-n}, x_{-n+1}, \ldots, x_{-1}, x_0,
x_{1}, \ldots, x_{n-1}, x_{n}$, starting from the beginning of the
first segment at $x_{-n}$ where the neutrino enters the Earth and
following the neutrino path until it reaches the detector at
$x_{n}$. For convenience, we have define a point $x_0$ at the middle
point of the trajectory, which artificially splits in two equal
parts the segment of the trajectory in the ($N-n+1$)-th shell. In
this way, there are two segments with equal length in each shell: in
the ($N-n+k$)-th shell, with $k=1, \ldots, n$, the two segments have
boundaries $(x_{-k}, x_{-k+1})$ and $(x_{k-1}, x_{k})$.

In each segment of the trajectory the electron fraction
$Y_e$, the CC potential $\Vcc$ and the induced
parameters $\xi$, $\varphi$, $\omega$, $V$ and $\dM$ are constant.
Due to the radial symmetry of the Earth, we have $X_k
= X_{-k}$ for $X=\xi, \varphi, \omega, V, \dM$.
Since the matrix elements of the non-adiabatic Hamiltonian
$\HMna$ of Eq.~(\ref{eq:HM}) vanish in each segment
and $\dM$ is constant,
we obtain
\begin{align}
  \SM(x_k,x_{k-1}) &\simeq {\rm exp}\large\{-i \intx \HMad \dx\large\}\nonumber\\
  &={\rm diag}\{e^{i \dM (x_k-x_{k-1})}, e^{-i \dM (x_k-x_{k-1})}\}
  \equiv \SM(x_k-x_{k-1})\quad, \label{eq:slab_SMad}
\end{align}
and accordingly
\begin{align}
\SV(x_k,x_{k-1})=\left[ \W \SM(\Delta x_k) \WD \right]_{(k)}\equiv
\SV_k \quad,\label{eq:slab_SVad}
\end{align}
where $\Delta x_k=x_k-x_{k-1}$ and the subscript $(k)$ indicates the ($N-n+k$)-th
shell.
Since $ \PsiV_2(x_k) = \SV_k \PsiV_2(x_{k-1}) $,
the evolution inside the Earth is given recursively by
\begin{align}
  \PsiV_2(x_n) \simeq \SVslab \PsiV_2(x_{-n})
  \label{eq:stat-endV}
  \quad,
\end{align}
with
\begin{align}
  \SVslab &= \SV_n \SV_{n-1} \ldots \SV_1 \SV_{-1} \ldots \SV_{-(n-1)} \SV_{-n}
  \nonumber{}\\
  &= \SV_n \SV_{n-1} \ldots \SV_1 \SV_{1} \ldots \SV_{n-1} \SV_{n}
  \quad. \label{eq:SV_slab}
\end{align}
Note that there are two $\SV_1$ in Eq.~(\ref{eq:SV_slab}) because we
have divided the innermost segment into two connected ones with the
central point $x_0$. By using
Eqs.~(\ref{eq:formal_ptb}),~(\ref{eq:stat-endV}) and
(\ref{eq:SV_slab}), we can calculate the transition probability
$\Ptb$ in a self-consistent way. This method is usually adopted in the
data analysis of solar neutrino experiments, because it is much more
rapid and efficient than the perturbation method.

Employing the perturbative approximation for the effective potential
$V$, we can further simplify the transition probability $\Ptb$ to
the first order of $V_k$ as
\begin{align}
  \Ptb(x_n) &\simeq \Ptb[V] + \cos^2\chi_\beta \sin2\theta_\beta \times
  \nonumber{} \\
  & \sum^n_{k=1} \frac{V_k}{\delta} \sin2\xi_k \sin( \delta L_n + {\Phi_{\beta}}_{(k)} ) \left( \sin \delta L_k - \sin \delta L_{k-1} \right)
  \quad,\label{eq:Ptb_slab_pert}
\end{align}
where $L_k$ is the length of the trajectory inside the outer boundary of the ($N-n+k$)-th shell:
\begin{align}
  L_k
  &=
  x_k - x_{-k}
  \quad. \label{eq:x_k}
\end{align}
The derivation of this expression is described in details in
Appendix~\ref{sec:detailed_slab}. We also derive in
Appendix~\ref{sec:3-flavor} the limit in the case of vanishing
active-sterile mixing of the expression (\ref{eq:Ptb_slab_pert}) for
the transition probability $\Pte$, in order to show that it agrees
with that derived in the case of standard three-neutrino mixing in
Ref.~\cite{liao}.

Finally, let us discuss the dependence
in $\Ptb$ and $R_{2\beta}$
on the mixing parameters. From Eqs.~(\ref{eq:Ptb_pert_V}) and
(\ref{eq:Ptb_slab_pert}), one can see that the regeneration factors
are explicitly proportional to the mixing parameters
$\cos^2\chi_\beta$ and $\sin2\theta_\beta$. Moreover, the
effective potential $V$, the effective mixing angle $\xi$ and the
oscillation wave number in matter $\dM$ take part in the
regeneration factors via the integration or summation of the
contributions in the whole trajectory inside the Earth. The most
distinct property in $\Ptb$ is the explicit dependence on the CP-violating phases
$\Phi_{\beta}$, which contribute to the oscillating phases.

\section{Numerical Discussion}
\label{sec:Numerical}

In the previous Section we have obtained in
Eqs.~(\ref{eq:Ptb_pert_V}), (\ref{eq:stat-endV}) and
(\ref{eq:Ptb_slab_pert}) three approximate analytical expressions
for the transition probability $\Ptb$. We will call them,
respectively, the perturbative approximation (PERT), the slab
approximation (SLAB) and the slab plus perturbative approximation
(SLAB$+$PERT). In this Section, we are going to test the accuracy of
these analytical approximations in the case of $3+1$ mixing and
discuss the main properties of the corresponding day-night
asymmetries.

First of all, in order to calculate the neutrino evolution inside the Earth
we need an accurate description of the density profile of
the Earth. In our calculations, we employ a simplified version of the
preliminary Earth reference model (PREM) \cite{PREM}, which contains
\cite{Lisi1997} five shells and uses the polynomial function
\begin{align}
  N_i(r)=\alpha_i + \beta_i\,r^2 + \gamma_i\,r^4
  \quad,
\end{align}
for the $i$-th shell ($1\leqslant i\leqslant 5$, where $i=1$ is the innermost
shell) to describe the Earth's density at the radial
distance $r$. The values of the coefficients are given in
Table~\ref{tb:Ne_earth}, which gives also the averaged
electron density $\overline{N}_{e\,i}$ in the $i$-th shell, which will be used in the slab
approximation. The electron fractions ${Y}_{e}$ are also provided
for the core and mantle regions. The cosine values of the nadir
angle $\kappa_{\rm N}$ corresponding to the radial boundaries
$[r_{i-1},\,r_i]$
of each shell are given in
the form
$[\cos\kappa_{\mathrm{N}\,i-1},\cos\kappa_{\mathrm{N}\,i}]$.
We employ the
fourth-order Runge-Kutta method for the numerical solution of the neutrino evolution equation
in the Earth,
using the density matrix formalism (see Appendix C of Ref.~\cite{CP} for a detailed introduction).
Finally, in Appendix \ref{sec:paramU} we give the
explicit parametrization of $U$ and the values of the
oscillation parameters used in our examples.

\begin{table}
\begin{center}
\scriptsize
\begin{tabular}{c|cccccccc}
\hline\hline $i$ & {\text{Shell}} & $[r_{i-1},\,r_i]$ &
$[\cos\kappa_{\mathrm{N}\,i-1},\cos\kappa_{\mathrm{N}\,i}]$ &
$\alpha_i$ & $\beta_i$ & $\gamma_i$ & $\overline{N}_{e\,i}$ &
${Y}_{e\,i} $
\\
\hline\hline
1 & Inner core      &     $[0,\,0.192]$    & $[1,\,0.98]$    &   6.099 & $-$4.119 &    0.000 & 6.048 & 0.466 \\
2 & Outer core      &   $[0.192,\,0.546]$  & $[0.98,\,0.84]$ &   5.803 & $-$3.653 & $-$1.086 & 5.209 & 0.466 \\
3 & Lower mantle    &   $[0.546,\,0.895]$  & $[0.84,\,0.45]$ &   3.156 & $-$1.459 &    0.280 & 2.468 & 0.494 \\
4 & Transition Zone &   $[0.895,\,0.937]$  & $[0.45,\,0.35]$ &$-$5.376 &   19.210 &$-$12.520 & 1.922 & 0.494 \\
5 & Upper mantle    &     $[0.937,\,1]$    & $[0.35,\,0]$    &  11.540 &$-$20.280 &   10.410 & 1.689 & 0.494 \\
\hline\hline
\end{tabular}
\end{center}
\caption{ \label{tb:Ne_earth} Descriptions of the simplified PREM
model with five shells. The shell names and the values of the
coefficients are quoted from Table 1 of Ref.~\cite{Lisi1997} (see
text for details).}
\end{table}

\begin{figure}
\begin{center}
\begin{tabular}{c}
\includegraphics*[width=0.8\textwidth]{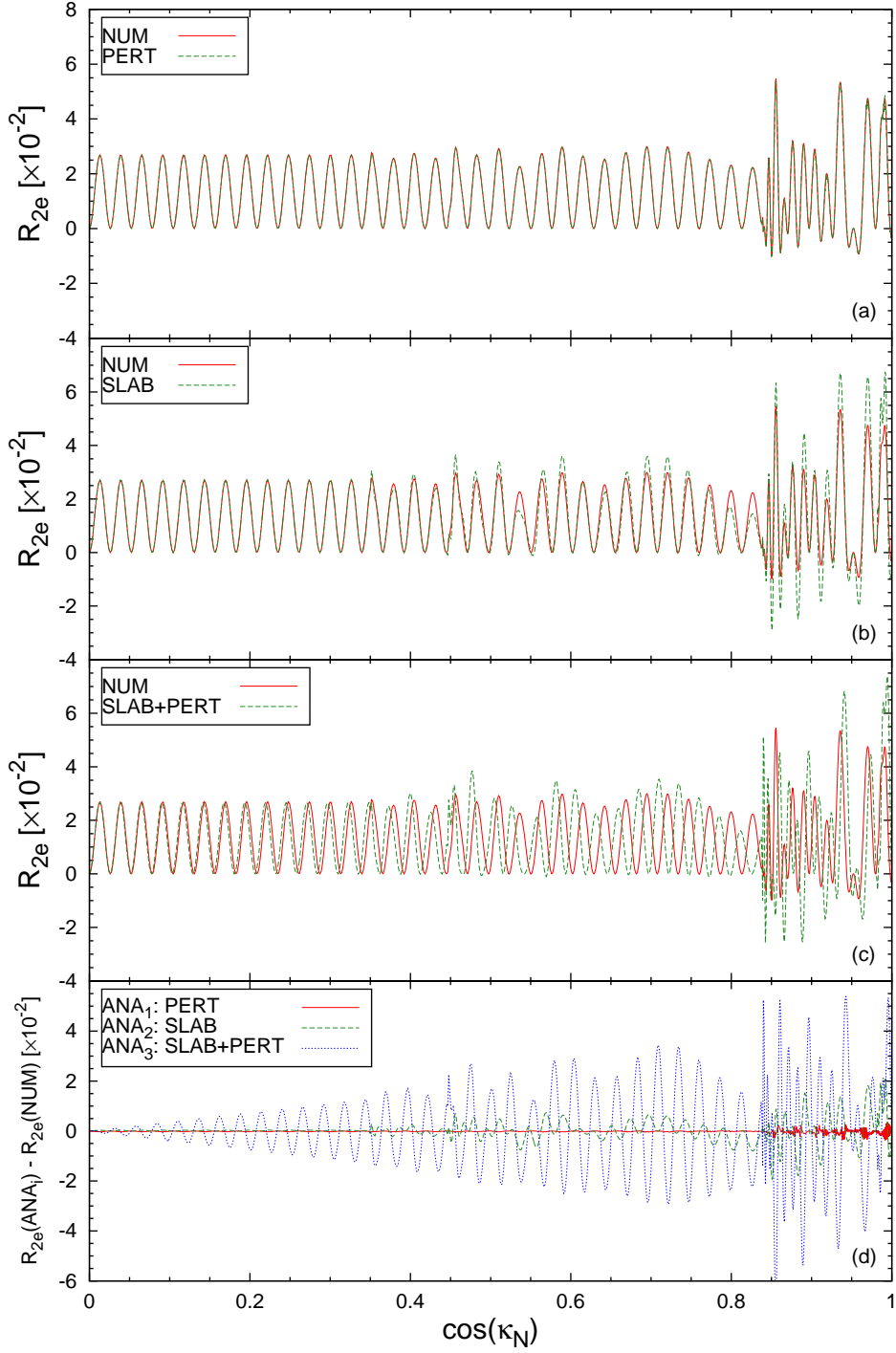} 
\end{tabular}
\end{center}
 \caption{\label{fig:Validation_fee}
Comparisons of the numerical calculation of the regeneration factor
$R_{2e}$ as the function of the nadir angle with the corresponding
analytical ones in the PERT, SLAB and SLAB+PERT approximations,
respectively. The lowermost panel is illustrated with the three
differences between the numerical and analytical results. The
neutrino energy is fixed to 10 MeV. All the oscillation parameters
are set to M1 and P1 in Eqs.~(\ref{M1}) and (\ref{P1}).}
\end{figure}
\begin{figure}
\begin{center}
\begin{tabular}{c}
\includegraphics*[width=0.8\textwidth]{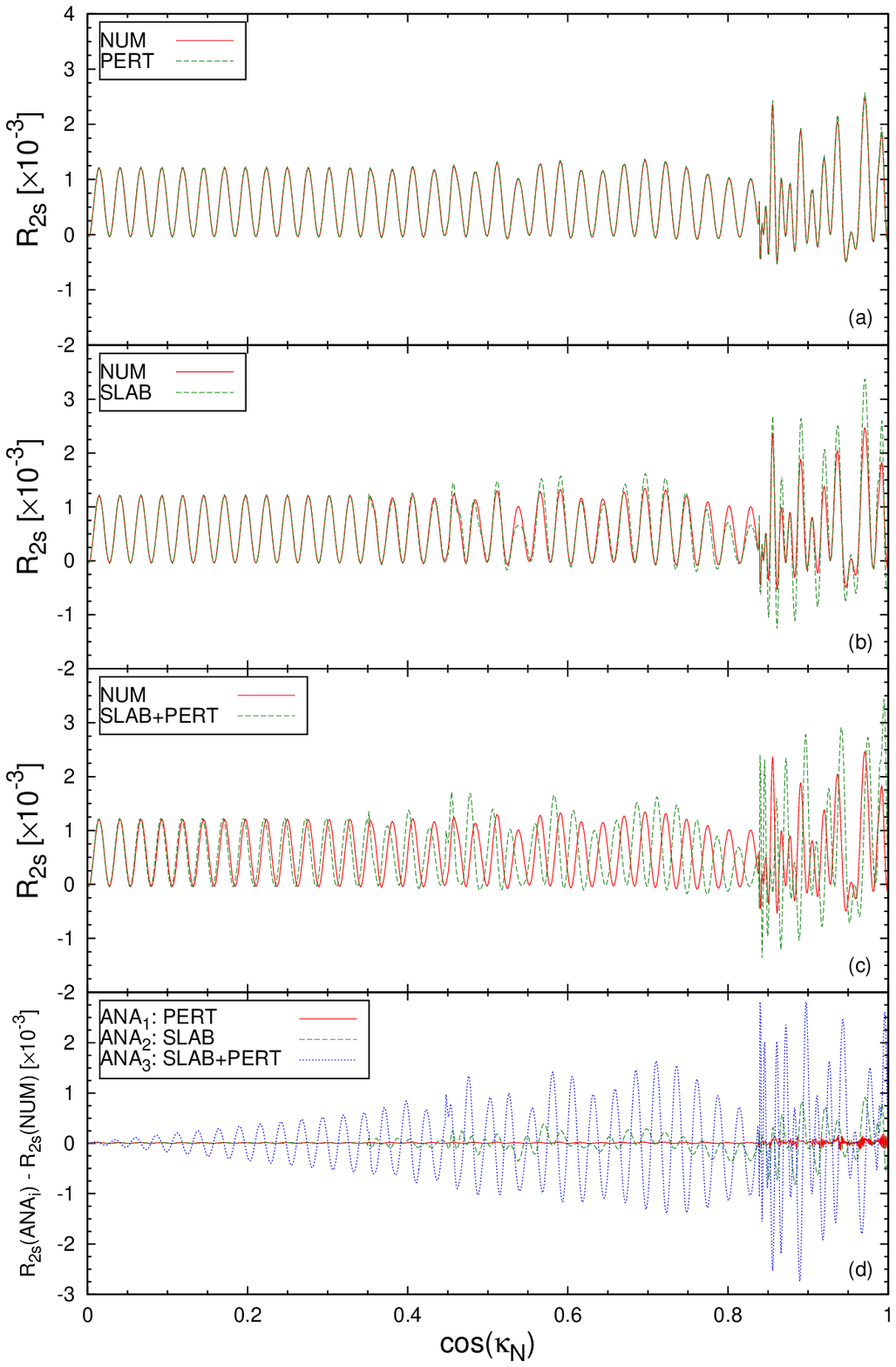}
\end{tabular}
\end{center}
 \caption{\label{fig:Validation_fes}
 The same as Figure \ref{fig:Validation_fee}, but for the regeneration factor $R_{2s}$.}
\end{figure}

In Figs.~\ref{fig:Validation_fee} and \ref{fig:Validation_fes}, we
check the accuracy of the three analytical approximations by
drawing the regeneration factors $R_{2\beta}$ (i.e., $R_{2e}$
and $R_{2s}$) as functions of the nadir angle, for a neutrino energy of $E=10\,\,{\rm
MeV}$. The three upper panels show the comparison of the result of each
analytical approximation
(PERT, SLAB and SLAB+PERT from top to bottom)
with that of the numerical calculation.
In the lowermost panels we have drawn the differences between the
analytical and numerical results as functions of the nadir
angle.

From Figs.~\ref{fig:Validation_fee} and \ref{fig:Validation_fes}, we
observe that the PERT approximation gives an accurate description of
the solar neutrino evolution inside the Earth, even for the neutrino
trajectories crossing the core (i.e. for $\cos\kappa_{\rm
N}\gtrsim0.84$). One can also see that the SLAB approximation gives
the correct frequencies of the oscillatory behavior of $R_{2e}$ and
$R_{2s}$, but there are significant differences in the oscillating
amplitudes with respect to those obtained with the numerical
calculation for neutrinos crossing the core. However, as we will
show later, these discrepancies are acceptable when the regeneration
factors are averaged over the relevant ranges of the energy and/or
nadir angle. Finally, one can notice that the accuracy of the
SLAB+PERT approximation is low, both for the frequencies and the
amplitudes of the oscillatory regeneration factors. Therefore, the
SLAB+PERT approximation can be used only for a qualitative analysis
and is not suitable for any realistic calculations.

The PERT approximation is the most accurate one, but it turns out to
be very time-consuming because of the two-dimensional numerical
integration in Eq.~(\ref{eq:Ptb_pert_V}). Hence, in the data
analysis of solar neutrino experiments it is convenient to employ
the SLAB approximation, which gives a rapid and approximately
accurate description for the terrestrial matter effect. In the
following discussion, we use the SLAB approximation as our default
choice.

The regeneration factors and the day-night asymmetries depend on the
nadir angle $\kappa_{\rm N}$ of the incoming solar neutrinos, and
thus on the latitude of the experimental site. With the aim of
illustrating the dependence on the site latitude, we define the
annual averaged quantity
\begin{align}
  \overline{Q}(E) &= \frac{ \int^{\kappa^{\rm max}_{\rm N}}_{\kappa^{\rm min}_{\rm N}} W(\kappa_{\rm N})
  \,Q(E,\kappa_{\rm N}) \, {\rm d}\kappa_{\rm N}}
  {\int^{\kappa^{\rm max}_{\rm N}}_{\kappa^{\rm min}_{\rm N}}  W(\kappa_{\rm N})\,{\rm d}\kappa_{\rm N}}
  \,,\label{eq:AverQ_E}
\end{align}
where $Q=R_{2\beta} \text{ or } D_{e\beta}$, and the weight function
$W(\kappa_{\rm N})$ represents the solar exposure of the trajectory
and depends on the latitude of the experimental site. $\kappa^{\rm
max}_{\rm N}$ and $\kappa^{\rm min}_{\rm N}$ are the possible
maximal and minimal values of the nadir angle at a certain latitude
(see Table II in Ref.~{\cite{Lisi1997}}). In our calculation, we use
the form of $W(\kappa_{\rm N})$ defined in Appendix C of
Ref.~{\cite{Lisi1997}}, in order to illustrate the effect of the
average over the nadir angle.

\begin{table}
\begin{center}
\begin{tabular}{|c|cc|c|cc|}
\hline\hline  & \text{Experimental Site} & Latitude & & \text{Experimental Site} & Latitude\\
\hline\hline
1 & Kamioka & $36.42^{\circ}$N  &4 & Kaiping  & $22.15^{\circ}$N\\
2 & Gran Sasso  & $42.46^{\circ}$N &5 & Pyhasalmi  & $63.66^{\circ}$N \\
3 & Sudbury & $46.47^{\circ}$N &6 & South Pole   & $90^{\circ}$S \\
\hline\hline
\end{tabular}
\end{center}
\caption{ \label{tb:3site} Latitudes of six experimental sites for
three ongoing experiments (left) and three planned experiments
(right).}
\end{table}
\begin{figure}
\begin{center}
\begin{tabular}{c}
\includegraphics*[bb=55 55 400 295, width=0.8\textwidth]{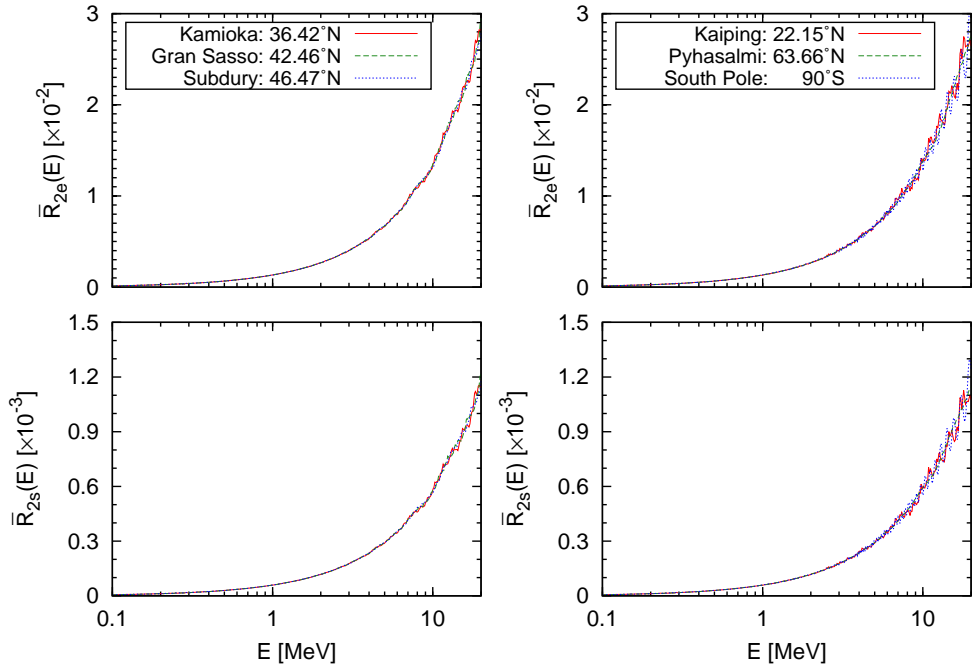}
\end{tabular}
\end{center}
 \caption{\label{fig:AverR2x_E(Latitude)}
Annual averages of the regeneration factors over the nadir angle
$\kappa_{\rm N}$. The experimental site latitudes are selected for
six typical examples in Table \ref{tb:3site}. All the oscillation
parameters are set to M1 and P1 in Eqs.~(\ref{M1}) and (\ref{P1}).}
\end{figure}
In the following we consider for illustration six experimental
sites: three with ongoing neutrino experiments (Kamioka in Japan,
Gran Sasso in Italy, and Sudbury in Canada) and three with planned
neutrino experiments (Kaiping \cite{dyb2} in China, Pyhasalmi
\cite{LENA} in Finland, and the South Pole \cite{icesolar}). Their
latitudes are listed in Table \ref{tb:3site}. The left and right
panels in Fig.~\ref{fig:AverR2x_E(Latitude)} show the behavior as a
function of the neutrino energy $E$ of the annual averages during
the nighttime of the regeneration factors $R_{2e}$ and $R_{2s}$
calculated with the SLAB approximation for the six typical examples
in Table~\ref{tb:3site}. The oscillation parameters are set to M1
and P1 in Eqs.~(\ref{M1}) and (\ref{P1}). One can see that the
annual averages of the regeneration factors increase almost
monotonously with the neutrino energy, except for tiny oscillatory
variations.

The behavior of the averaged regeneration factors in
Fig.~\ref{fig:AverR2x_E(Latitude)} can be understood with the help
of Eq.~(\ref{eq:Ptb_slab_pert}). The magnitudes of the regeneration
factors are proportional to the neutrino energy via the amplitude
term ${V_k}/{\delta}$, which contributes dominantly to the shape of
the curves and determines the sizes of the regeneration factors. The
residual oscillatory behavior is due to the summation of the
functions $\sin( \delta L_n + {\Phi_{\beta}}_{(k)})(\sin \delta L_k
- \sin \delta L_{k-1})$ for $n\leqslant5$ and $1\leqslant k
\leqslant n$, which oscillate as functions of the trajectory length,
which is determined by the nadir angle. The averaging over the nadir
angle reduces the size of the oscillation amplitudes. The size and
pattern of the oscillatory behavior are different for different site
latitudes, thus their effects are more significant in the right
panels of Fig.~\ref{fig:AverR2x_E(Latitude)}, where the experimental
sites range from the tropic area to the South Pole. In the following
discussion, we consider only the Kamioka site for simplicity.

As we have illustrated in Figs.~\ref{fig:Validation_fee} and
\ref{fig:Validation_fes}, the accuracy of the SLAB approximation is
low for neutrinos crossing the core of the Earth when one considers
precise values of the energy and nadir angle. Now we check the
validity of the SLAB approximation in a more realistic case, in which
the regeneration factors are averaged over bins of the nadir angle
$\kappa_{\rm N}$ and the neutrino energy $E$. The annual average of
the regeneration factor in the ($i$, $j$) bin is defined as
\begin{align}
  \overline{R}_{2\beta} &= \frac{ \int^{E^{\rm max}_j}_{E^{\rm min}_j}
  \int^{\kappa^{\rm max}_{{\rm N}\,i}}_{\kappa^{\rm min}_{{\rm N}\,i}}
  W(\kappa_{\rm N})W^{\prime}(E)
  \,{R}_{2\beta}(E,\kappa_{\rm N}) \, {\rm d}\kappa_{\rm N}\,{\rm d}E}
  {\int^{\kappa^{\rm max}_{{\rm N}\,i}}_{\kappa^{\rm min}_{{\rm N}\,i}}
  W(\kappa_{\rm N})\,{\rm d}\kappa_{\rm N}\times
  \int^{E^{\rm max}_j}_{E^{\rm min}_j} W^{\prime}(E)\,{\rm d}E}
  \,,\label{eq:AverQ_bin}
\end{align}
where $W(\kappa_{\rm N})$ and $W^{\prime}(E)$ are the weight
functions of the nadir angle and neutrino energy.
We use the expression of $W(\kappa_{\rm N})$ given in Appendix C of Ref.~{\cite{Lisi1997}}.
The weight function $W^{\prime}(E)$ depends on the energy-dependent
flux spectrum, the detection cross section and
the experimental efficiency.
However, for simplicity
we neglect the effect of $W^{\prime}(E)$, since
the energy bin-size is small compared to the variations of
$W^{\prime}(E)$.

\begin{figure}
\begin{center}
\begin{tabular}{c}
\includegraphics*[width=0.9\textwidth]{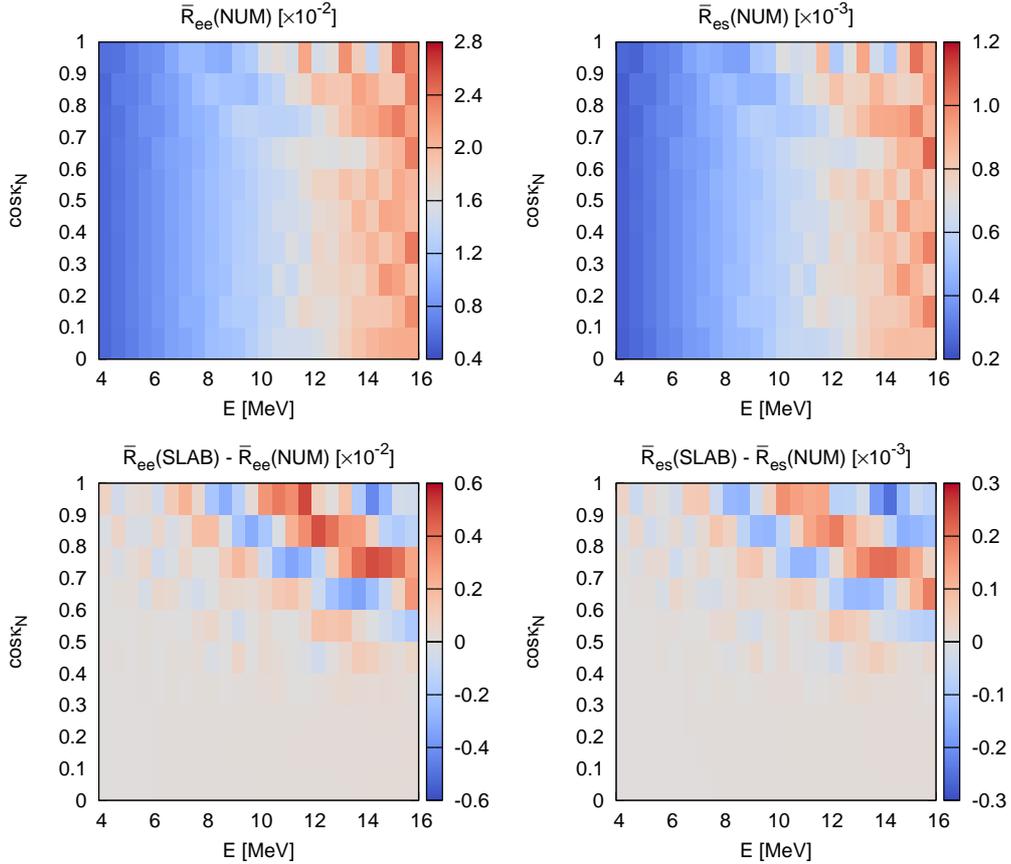}
\end{tabular}
\end{center}
 \caption{\label{fig:AverR2x_Bin}
Binned average of the numerical (NUM) calculations (upper panels) of
the regeneration factors $R_{2e}$ (left) and $R_{2s}$ (right) as the
functions of the neutrino energy and the nadir angle. The lower
panels illustrate the differences between the NUM and SLAB
calculations. All the oscillation parameters are set to M1 and P1 in
Eqs.~(\ref{M1}) and (\ref{P1}).}
\end{figure}
In the upper panels of Fig.~\ref{fig:AverR2x_Bin}, we show the
binned average of the numerical (NUM) calculations of the
regeneration factors $R_{2e}$ (left) and $R_{2s}$ (right) as
functions of the neutrino energy and nadir angle. The lower panels
illustrate the corresponding differences between the NUM and SLAB
calculations. All the oscillation parameters are set to M1 and P1 in
Eqs.~(\ref{M1}) and (\ref{P1}), and the latitude for the Kamioka
site is assumed. In our calculation, we choose equal-sized bins for
the neutrino energy ($\Delta E= 0.5 $ MeV) and the cosine of the
nadir angle ($\Delta\cos \kappa_{\rm N}= 0.1$). From the lower
panels of Fig.~\ref{fig:AverR2x_Bin}, we can observe that the
discrepancy between the NUM and SLAB calculations of the
regeneration factors is visible only for high neutrino energies and
small nadir angles. The accuracy of the SLAB calculation of the
binned average is excellent in most of the parameter space, and is
better than $20\%$ in the worst case. Thus, we conclude that the
SLAB approximation is reliable for practical applications.

\begin{figure}
\begin{center}
\begin{tabular}{c}
\includegraphics*[width=0.8\textwidth]{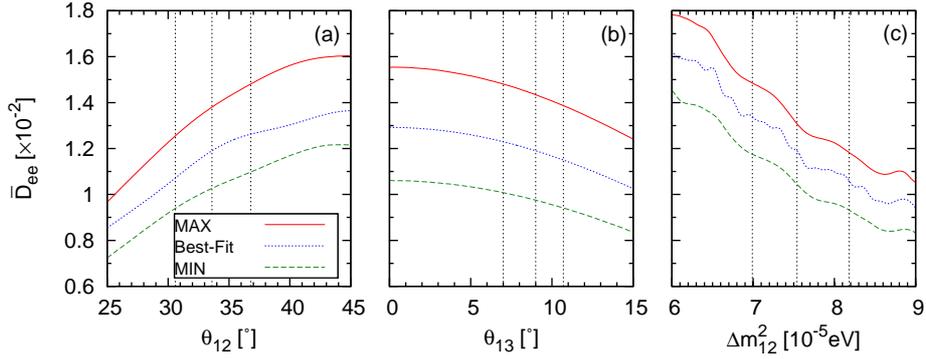}
\end{tabular}
\end{center}
\caption{\label{fig-Dee_3fParam} Dependence of $\averD[e]$ on the
oscillation parameters $\theta_{12}$ (left), $\theta_{13}$ (middle)
and $\Delta m^2_{21}$ (right) in the three-neutrino mixing scheme
for $E=10$ MeV and zero CP-violating phase. The solid and dashed
lines in each panel correspond to the maximal (MAX) and minimal (MIN)
values of $\averD[e]$ when the other parameters are scanned in the
full $\pm3\sigma$ ranges \cite{fogli}. The short-dashed line corresponds
to the best-fit values of these parameters. The three
vertical lines in each panel correspond to the best-fit and
$\pm3\sigma$ values of the parameter in abscissa.}
\end{figure}
Next, we are going to explore the properties of the annual averages
$\averD[e]$ and $\averD[s]$ defined in
Eq.~(\ref{eq:AverQ_E}) of the day-night asymmetries of solar
neutrino oscillations defined in Eq.~(\ref{eq:Debeta}).
Although the regeneration factors $\averR(E)$
are always positive in the whole energy range, $\averD$ can be both
positive or negative, depending on the sign of the parameter
$\cos2\Theta^0_{e}$. From the energy dependence of
$\cos2\Theta^0_{e}$ in Fig.~5 of Ref.\cite{CP}, we know that the
sign of $\cos2\Theta^0_{e}$ flips at $E\simeq2\,$MeV and thus the
sign of $\averD$ changes accordingly. For neutrino energies lower
than about $2\,$MeV, both $\averD[e]$ and $\averD[s]$ are negative and
their magnitudes are of the order of $10^{-4}$, which is too small
for the detection ability of ongoing and near-future solar neutrino
experiments. For neutrino energy greater than about $2\,$MeV, the
day-night asymmetries increase monotonously as the energy grows,
giving hope for their detection.

To illustrate the different contributions of the oscillation
parameters to the day-night asymmetries, we first show in Figure
\ref{fig-Dee_3fParam} the dependence of $\averD[e]$ on the relevant
oscillation parameters $\theta_{12}$ (left), $\theta_{13}$ (middle)
and $\Delta m^2_{21}$ (right) in the three-neutrino mixing scheme
for $E=10$ MeV. The solid and dashed lines in each panel correspond
to the maximal (MAX) and minimal (MIN) values of $\averD[e]$ when
the other parameters (e.g. $\theta_{12}$ and $\theta_{13}$ in the
right panel) are scanned in the full $\pm3\sigma$ ranges
\cite{fogli}. The short-dashed line corresponds to the best-fit
values of these parameters. The three vertical lines in each panel
correspond to the best-fit and $\pm3\sigma$ values of the parameter
in abscissa. One can see that the magnitude of the day-night
asymmetry is of the order of $10^{-2}$. Moreover, one can see that
$\averD[e]$ increases with $\theta_{12}$, but decreases as
$\theta_{13}$ increases. This can be explained with the help of
Eqs.~(\ref{eq:P2e_pert_3f}) and (\ref{eq:P2e_slab_3f}), which show
that the day-night asymmetry is proportional to
$\cos^4\theta_{13}\sin^2 2 \theta_{12}$. Considering the role of
$\Delta m^2_{21}$, higher $\Delta m^2_{21}$ induce lower values of
$\averD[e]$, which is consistent with the fact that the asymmetry is
proportional to $1/\delta = 4E/\Delta m^2_{21}$ in
Eq.~(\ref{eq:P2e_slab_3f}).

\begin{figure}
\begin{center}
\begin{tabular}{c}
\includegraphics*[width=0.9\textwidth]{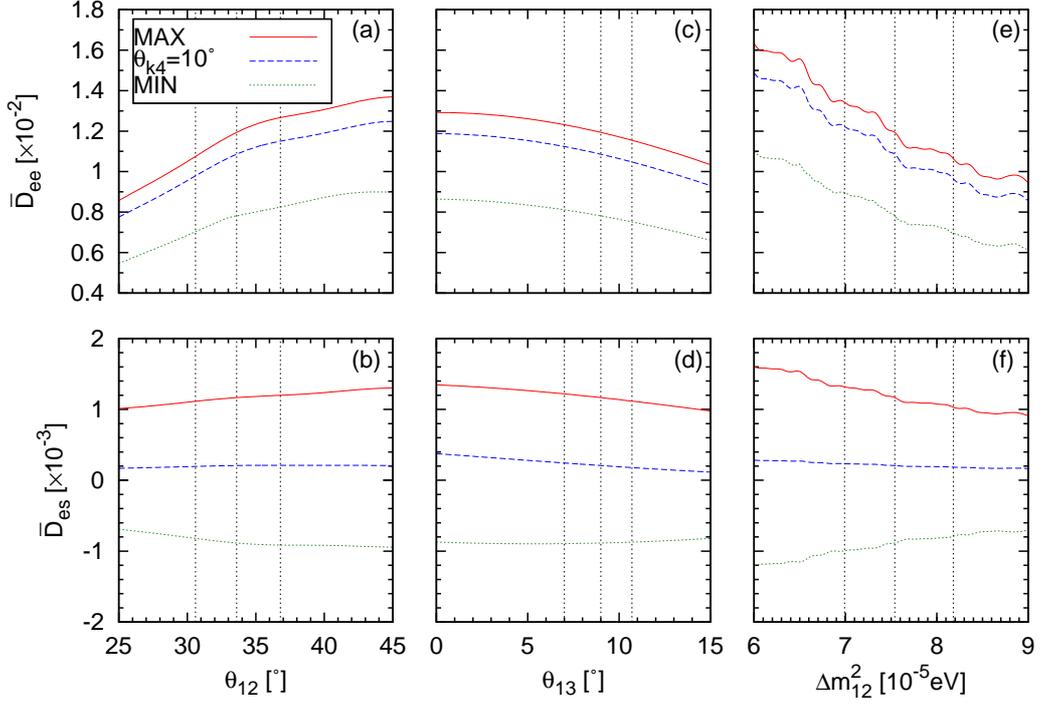}
\end{tabular}
\end{center}
\caption{\label{fig-Dex_4fParam(active)} Dependence of
$\averD[e]$(upper panels) and $\averD[s]$ (lower panels) on the
active neutrino oscillation parameters $\theta_{12}$ (left),
$\theta_{13}$ (middle) and $\Delta m^2_{21}$ (right) for $E=10$ MeV
and zero CP-violating phases. The solid and dashed lines correspond
to the MAX and MIN values of $\averD[e]$ or $\averD[s]$ when the
active-sterile mixing angles $\theta_{k4}$ are scanned in the interval
$[0^{\circ}, 20^{\circ}]$. The short-dashed lines correspond to
$\theta_{k4}=10^\circ$. The three vertical lines correspond to
the best-fit and $\pm3\sigma$ values of the parameter in abscissa.
In each panel, the values of the other active neutrino oscillation
parameters (e.g. $\theta_{13}$ and $\Delta m^2_{21}$ in the left
panels) are fixed to their best-fit values \cite{fogli}.}
\end{figure}
\begin{figure}
\begin{center}
\begin{tabular}{c}
\includegraphics*[width=0.9\textwidth]{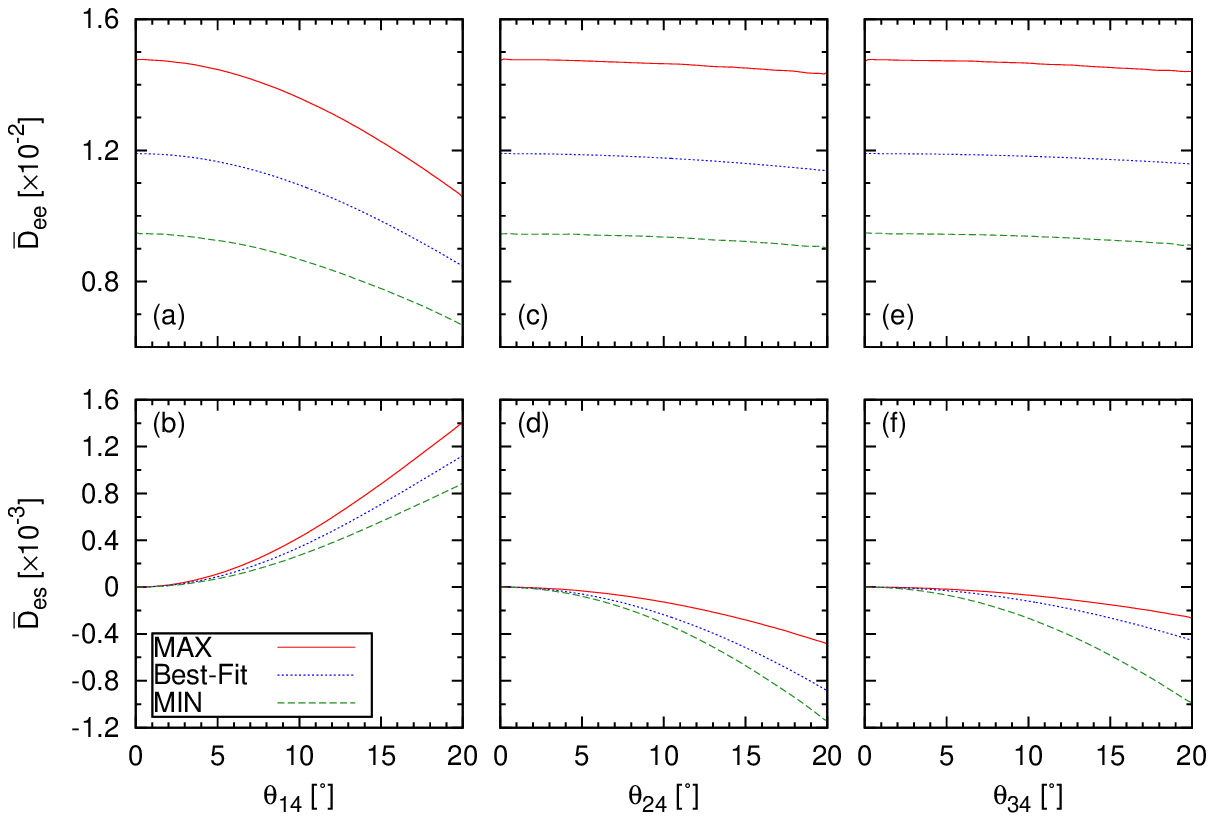}
\end{tabular}
\end{center}
 \caption{\label{fig-Dex_4fParam(sterile)}
Dependence of $\averD[e]$ (upper panels) and $\averD[s]$ (lower
panels) on the active-sterile neutrino mixing parameters
$\theta_{14}$ (left), $\theta_{24}$ (middle) and $\theta_{34}$
(right) for $E=10$ MeV and zero CP-violating phases. The solid and
dashed lines in each panel correspond to the MAX and MIN values of
$\averD[e]$ or $\averD[s]$ when the three active mixing angles are
scanned in the full $\pm3\sigma$ ranges \cite{fogli}. The
short-dashed line corresponds to the best-fit values of these
parameters. In each panel the values of the other active-sterile
mixing angles (e.g. $\theta_{14}$ and $\theta_{24}$ in the right
panels) are fixed to be zero.}
\end{figure}
Considering now the 3+1 mixing scheme, in
Fig.~\ref{fig-Dex_4fParam(active)} we illustrate the dependence of
$\averD[e]$(upper panels) and $\averD[s]$ (lower panels) on the
active neutrino oscillation parameters $\theta_{12}$ (left),
$\theta_{13}$ (middle) and $\Delta m^2_{21}$ (right) for $E=10$ MeV
and zero CP-violating phases. The solid and dashed lines correspond
to the MAX and MIN values of $\averD[e]$ or $\averD[s]$ when the
active-sterile mixing angles $\theta_{k4}$ are scanned in the interval
$[0^{\circ}, 20^{\circ}]$. The short-dashed lines correspond to
$\theta_{k4}=10^\circ$. The three vertical lines correspond to
the best-fit and $\pm3\sigma$ values of the parameter in abscissa.
In each panel, the values of the other active neutrino oscillation
parameters (e.g. $\theta_{13}$ and $\Delta m^2_{21}$ in the left
panels) are fixed to their best-fit values \cite{fogli}. Similar
illustrations are given in Fig.~\ref{fig-Dex_4fParam(sterile)} for
the active-sterile mixing parameters $\theta_{14}$ (left),
$\theta_{24}$ (middle) and $\theta_{34}$ (right). The
solid and dashed lines in each panel correspond to the MAX and MIN
values of $\averD[e]$ or $\averD[s]$ when the three active mixing
angles are scanned in the full $\pm3\sigma$ ranges \cite{fogli}. The
short-dashed line corresponds to the best-fit values of these
parameters. In each panel the values of the other active-sterile
mixing angles (e.g. $\theta_{14}$ and $\theta_{24}$ in the right
panels) are fixed to be zero.

From the upper panels of Fig.~\ref{fig-Dex_4fParam(active)},
one can see that the
dependence of $\averD[e]$ on $\theta_{12}$, $\theta_{13}$ and
$\Delta m^2_{21}$ is similar to that in the case of three neutrino
mixing\footnote{
In the specific parametrization
of $U$ in Eq.~(\ref{eq:Uparam}), the day-night asymmetries
depend also on the mixing angle $\theta_{23}$ due the
non-commutativity of the matrices, but the
dependence is negligible because of the smallness of the
active-sterile mixing.
}, but the contributions of active-sterile mixing
reduce the size of $\averD[e]$. Moreover, we can observe from
the upper panels of Fig.~\ref{fig-Dex_4fParam(sterile)} that
this suppression is mainly due to $\theta_{14}$.
This is explained by the presence of the factor
$\cos^2\chi_e = \cos^2\theta_{13} \cos^2\theta_{14}$ in Eq.~(\ref{eq:Debeta}),
whereas
the smallness of the effects of $\theta_{24}$ and $\theta_{34}$
on $\averD[e]$ is due to the fact that
they only contribute indirectly through the parameters
$\Theta^0_{e}$, $V$ and $\xi$.

In the lower panels of
Figs.~\ref{fig-Dex_4fParam(active)}
and~\ref{fig-Dex_4fParam(sterile)}, the magnitude of $\averD[s]$
grows as the active-sterile mixing becomes larger,
but the sign depends on the mixing angle:
it is positive for the contribution of
$\theta_{14}$
and negative for the contributions of
$\theta_{24}$ and $\theta_{34}$.
Therefore,
there are more possibilities to measure $\averD[s]$
if $\theta_{14}$
is much larger than
$\theta_{24}$ and $\theta_{34}$
or vice versa.
In any case,
high-precision experiments are needed, because the maximum value of
$\averD[s]$
is of the order of $10^{-3}$.

\begin{figure}
\begin{center}
\begin{tabular}{c}
\includegraphics*[bb=55 55 408 216, width=0.8\textwidth]{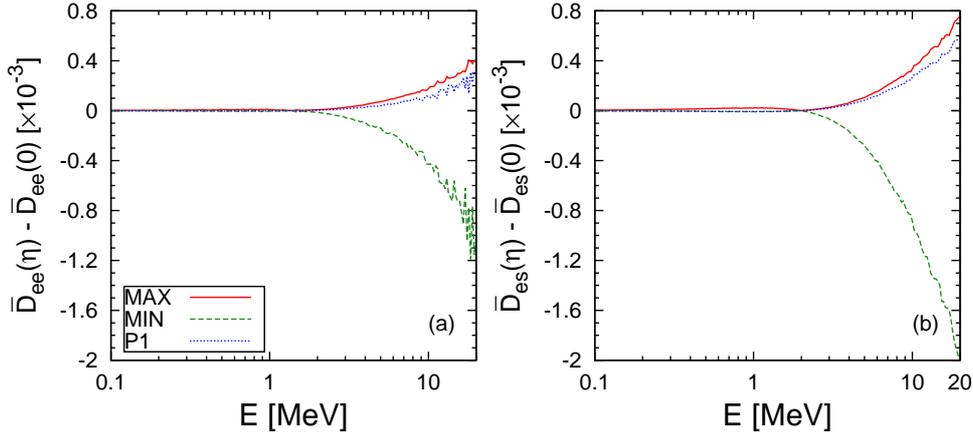}
\end{tabular}
\end{center}
  \caption{\label{fig:CPphase}
Energy spectra of the differences $\averD(\eta) - \averD(0)$ of the
annually averaged day-night asymmetry $\averD(\eta)$ obtained with
non-zero CP-violating phases and $\averD(0)$ obtained with vanishing
CP-violating phases. The mass and mixing parameters are set to M1 in
Eq.~(\ref{M1}). The three CP-violating phases $\eta_{13}$,
$\eta_{24}$, $\eta_{34}$ have been randomly scanned in the full
parameter space to produce the MAX and MIN boundary curves. The
short-dashed curve corresponds to the set of phases P1 in
Eq.~(\ref{P1}).}
\end{figure}
Finally, it is interesting to investigate the effects of the three
CP-violating phases on the day-night asymmetries $\averD[e]$ and
$\averD[s]$. The left (right) panel of Figure~\ref{fig:CPphase}
shows, as a function of the neutrino energy, the possible difference
between the annually averaged day-night asymmetry $\averD[e](\eta)$
($\averD[s](\eta)$) obtained with non-zero CP-violating phases
$\eta_{13}$, $\eta_{24}$, $\eta_{34}$ and $\averD[e](0)$
($\averD[s](0)$) obtained with vanishing CP-violating phases. The
MAX and MIN curves show, respectively, the maximal and minimal
values of the difference which have been obtained with a random scan
of all values of the CP-violating phases. Note that the values of
the phases in each of these two curves may be different at different
energies. From Fig.~\ref{fig:CPphase}, we can learn that the effects
of the CP-violating phases are negligible in the low energy region
and start to emerge at about 2 MeV. Above this energy, the effects
of the unknown CP-violating phases increase with the energy,
reaching a maximum level of the order of $10^{-3}$ for $E \gtrsim
10\,$ MeV. Therefore, the relative variations of the day-night
asymmetries due to the CP-violating phases can reach about $10\%$
for $\averD[e]$ and may be as large as $100\%$ for $\averD[s]$ in
the active-sterile transition. This conclusion is consistent with
those on the oscillation probabilities in the daytime \cite{CP}.

\section{Conclusion}
\label{sec:Conclusion}

In this paper, we have discussed solar neutrino oscillations
inside the Earth and
we have derived the
regeneration factors (\ref{eq:R2beta})
and the day-night asymmetries (\ref{eq:Debeta})
of solar neutrino oscillations in a general scheme of
3+$N_{s}$ neutrino mixing, without any constraint on the neutrino
mixing elements, assuming only a realistic hierarchy on the
mass-squared differences.
We have discussed two approximate analytical solutions of the
neutrino evolution inside the Earth,
the perturbative approximation and the slab approximation,
which allow us to calculate
the regeneration factors and the day-night asymmetries
with different accuracy levels and
computational burdens.

In Section~\ref{sec:Numerical} we have presented several examples of
numerical calculations of solar neutrino active-sterile oscillations inside the
Earth in the 3+1 neutrino mixing scheme.
By using a simplified version of the
PREM description of the Earth density profile, we tested the
analytical approximations and learned that the
perturbative approximation can give the most accurate description of
the neutrino evolution inside the Earth and the slab approximation
gives less accurate but more efficient calculations.

We have shown that
the annual averages
$\averD[e]$ and $\averD[s]$ of the
$\nu_{e}\to\nu_{e}$
and
$\nu_{e}\to\nu_{s}$
day-night asymmetries are insensitive to the latitude differences of the
experimental sites and increase with the neutrino energy.
The magnitudes of
$\averD[e]$ and $\averD[s]$
can reach, respectively, a size of the order of $10^{-2}$ and $10^{-3}$
for high-energy solar neutrinos.
The three active-sterile mixing angles $\theta_{14}$, $\theta_{24}$ and
$\theta_{34}$ have a suppression effect on $\averD[e]$
(mostly $\theta_{14}$).
Studying the dependence of $\averD[s]$
on the active-sterile mixing angles,
we found that
the active-sterile mixing angle $\theta_{14}$
generates a positive contribution to $\averD[s]$,
whereas the contribution of $\theta_{24}$ and $\theta_{34}$
is negative.
Therefore,
there are more possibilities to measure $\averD[s]$
if $\theta_{14}$
is much larger than
$\theta_{24}$ and $\theta_{34}$
or vice versa.
Finally, we have shown that the variations of the
annual averages $\averD[e]$ and $\averD[s]$
induced by the unknown CP-violating phases can be at
the level of $10^{-3}$ in the high energy region. Therefore, it
might be possible to observe the effects of active-sterile
mixing and that of the CP-violating phases in the day-night asymmetries
in future high-precision solar neutrino experiments.

\vspace{0.5cm}
\section*{Acknowledgment}
\label{sec:acknowledgment}

H.W. L. would like to thank Prof.~Peng-Fei Zhang for his continuous
encouragement and financial support. The work of H.W. L. is
supported in part by the National Natural Science Foundation of
China under Grant No. 11265006. The work of Y. F. L. is supported in
part by the National Natural Science Foundation of China under Grant
No. 11135009.

\newpage
\appendix{}

\section{Details of the Analytical Calculation}
\label{sec:details}

Since the $S$-matrix of the neutrino evolution in the Earth defined
in Eq.~(\ref{eq:S_PsiVorM}) is a $2 \times 2$ unitary matrix, we can
rewrite $\SV$ using the Pauli matrices,
\begin{align}
  \SV &=  Q_0 \cdot \mathbf{1} - i \vec\sigma \cdot \vec{Q}
  \nonumber{}\\
  &= Q_0 \cdot \mathbf{1} - i( Q_1 \cdot \sigma_1 + Q_2 \cdot \sigma_2 + Q_3 \cdot \sigma_3 )
  \quad.\label{eq:expand_Pauli}
\end{align}
Then from Eqs.~(\ref{eq:formal_ptb}) and (\ref{eq:S_PsiVorM}), and
the initial condition, we have
\begin{align}
  \Ptb(x_f) &= \half \cos^2\chi_\beta \Big\{ 1 + \cos2\theta_\beta ( Q_1^2 + Q_2^2 -Q_3^2 - Q_0^2 )
  \nonumber{}\\
  &- 2 \sin2\theta_\beta \cos(\phi_{\beta 2} - \phi_{\beta 1}) (Q_0Q_2 + Q_1Q_3)
  \nonumber{}\\
  &- 2 \sin2\theta_\beta \sin(\phi_{\beta 2} - \phi_{\beta 1}) (Q_0Q_1 - Q_2Q_3) \Big\}
  \quad.\label{eq:ptb_Q}
\end{align}
All the following calculations are based on this expression.

\subsection{Perturbative Approximation}
\label{sec:detailed_pert}

Comparing Eqs.~(\ref{eq:SMpX}) and~(\ref{eq:expand_Pauli}), we can
get the expressions of $Q_i$ in the perturbative approximation as
\begin{align}
  Q_0 &= \cos\Delta - A \sin\Delta\,,
  \nonumber{}\\
  Q_1 &= \mRe(C) \cos\Delta - \mIm(C) \sin\Delta\,,
  \nonumber{}\\
  Q_2 &= - \mRe(C) \sin\Delta - \mIm(C) \cos\Delta\,,
  \nonumber{}\\
  Q_3 &= - \sin\Delta - A \cos\Delta
  \quad.
\end{align}
According to Eq.~(\ref{eq:ptb_Q}) and to the first order of $A$ and
$C$, we obtain
\begin{align}
  \Ptb(x_f) \simeq& \Ptb[V] + \cos^2\chi_\beta \sin2\theta_\beta \times
  \nonumber{}\\
  &\left[ \mRe(C) \sin( 2\Delta + \phi_{\beta1} - \phi_{\beta2} ) + \mIm(C) \cos( 2\Delta + \phi_{\beta1} - \phi_{\beta2} ) \right]
  \quad,
\end{align}
which leads to Eq.~(\ref{eq:Ptb_pert_C}).

The electron fraction $Y_e$ can be approximated as constant in the
Earth, so we can further simplify the corresponding term of $C$ as
\begin{align}
  &\mIm[ C e^{ i (2\Delta + \phi_{\beta1} - \phi_{\beta2}) } ] =  \intx - \dw \cos[2\Delta(x_f,x) + \Phi_{\beta} ] \dx
  \nonumber{}\\
  &= -\omega \cos[2\Delta(x_f,x) + \Phi_{\beta}] \Big|_{x_i}^{x_f}
  + \intx \omega \dM \sin[2\Delta(x_f,x) + \Phi_{\beta} ] \dx
  \quad,
\end{align}
where $\Phi_{\beta}=\phi_{\beta1} - \phi_{\beta2} + \varphi$.

Since $\omega(x_i)=0$, $\Delta(x_f,x_f)=0$ and to the first order of
$V$
\begin{align}
  \omega \simeq \frac{ \sin2\xi }{ 2\delta } V\quad,\quad\quad
  \dM \simeq \delta - \cos2\xi \cdot V
  \quad,
\end{align}
we obtain the perturbative approximation of $\Ptb(x_f)$
in Eq.~(\ref{eq:Ptb_pert_V}).

\subsection{Slab Approximation}
\label{sec:detailed_slab}

To calculate the transition probability $\Ptb$ in the slab
approximation, it is useful to rewrite $\SVslab$ and $\SV_k$
in Eq.~(\ref{eq:SV_slab}) in terms of the Pauli matrices.
First, for $\SV_k$ we have
\begin{align}
  \SV_k &= \left[ \W \SMad(\Delta x_k) \WD \right]_{(k)}
  \nonumber{}\\
  &= c_k \cdot \mathbf{1} - i \vec\sigma \cdot \vec{s_k}
  \quad,\label{eq:expand_Pauli_slab}
\end{align}
with
\begin{align}
  c_k &\equiv \cos\eta_k
  \quad,\\
  \vec s_k &\equiv \sin\eta_k ( \sin2\omega_k \cos\varphi_k, -\sin2\omega_k \sin\varphi_k, -\cos2\omega_k )
  \quad,\\
  \eta_k &\equiv \dMk \Delta x_k
  \quad.
\end{align}
Then, we can define
\begin{align}
\SVslabk &= c_k^{\;'} \cdot \mathbf{1} - i \vec\sigma \cdot \vec
s_k^{\;'}
\end{align}
with
\begin{align}
  \SVslabk &= \SV_k {\SVslab}_{(k-1)} \SV_k
  \quad,\\
  {\SVslab}_{(0)} &= \mathbf{1}
  \quad.
\end{align}
Therefore, we obtain
\begin{equation}
  \left\{
    \begin{array}{ll}
      c_k^{\;'} &= c_{k-1}^{\;'} (2c_k^2-1) - \vec s_{k-1}^{\;'} \cdot (2c_k\vec s_k)\quad,
      \\
      \vec s_k^{\;'} &= c_{k-1}^{\;'} (2c_k\vec{s}_k) - 2 (\vec{s}_{k-1}^{\;'} \cdot \vec s_k) \vec{s_k} + \vec{s}_{k-1}^{\;'}\quad,
      \\
      c_0^{\;'} &=1\quad,
      \\
      \vec s_0^{\;'} &= (0,0,0)\quad,
    \end{array}
  \right. \quad
\end{equation}
or, in matrix form,
\begin{align}
  (c_k^{\;'} , \vec s_k^{\;'}) &= (c_{k-1}^{\;'} , \vec s_{k-1}^{\;'}) T_k
  = (1,0,0,0) \prod_{i=1}^k T_i
  \quad.
\end{align}
Here $T_k$ is the $4 \times 4$ matrix
\begin{align}
  &T_k =
  \begin{pmatrix}
    2c_k^2-1 & 2c_k \vec s_k \\
    -2c_k \vec s_k^{\,\mathrm{T}} & \mathbf{1} - 2\vec s_k \otimes \vec s_k
  \end{pmatrix}
  =
  \nonumber\\
  &\footnotesize{
  \begin{pmatrix}
    \cos2\eta & \cos\varphi \sin2\eta \sin2\omega & - \sin\varphi \sin2\eta \sin2\omega & - \sin2\eta \cos2\omega
    \\
    - \cos\varphi \sin2\eta \sin2\omega & 1 - 2 \cos^2\varphi \sin^2\eta \sin^22\omega & \sin2\varphi \sin^2\eta \sin^22\omega & \cos\varphi \sin^2\eta \sin4\omega
    \\
    \sin\varphi \sin2\eta \sin2\omega & \sin2\varphi \sin^2\eta \sin^22\omega & 1 - 2 \sin^2\varphi \sin^2\eta \sin^22\omega & - \sin\varphi \sin^2\eta \sin4\omega
    \\
    \sin2\eta \cos2\omega & \cos\varphi \sin^2\eta \sin4\omega & - \sin\varphi \sin^2\eta \sin4\omega & 1 - 2 \cos^22\omega \sin^2\eta
  \end{pmatrix}_{(k)}}\,,
\end{align}
where the subscripts $(k)$ and $k$ denote the quantities in the ($N-n+k$)-th shell.
Finally, we obtain the following
compact expression for $\SVslab$ which allows to calculate the
evolution of the amplitudes in the Earth according to Eq.~(\ref{eq:stat-endV}):
\begin{align}
  \SVslab = ( \mathbf{1} , -i\vec\sigma ) \cdot ( c_n^{\;'} , \vec s_n^{\;'})
  = (\mathbf{1} , -i\vec\sigma ) \cdot \left[ (1,0,0,0) \prod_{i=1}^n T_i \right]
  \quad. \label{eq:SVslab_T}
\end{align}
This expression corresponds to that in Eq.~(\ref{eq:expand_Pauli})
with
\begin{equation}
Q_j = \left( \prod_{i=1}^n T_i \right)_{0j}
\,,
\quad
\mathrm{for}
\quad
j=0,1,2,3
\,,
\label{Qj}
\end{equation}
and allows to calculate $\Ptb(x_n)=\Ptb(x_f)$ through Eq.~(\ref{eq:ptb_Q}).

Let us now derive the
perturbative approximation in Eq.~(\ref{eq:Ptb_slab_pert}).
To the first order of $V_k$, we can approximate $T_k$ by
\begin{align}
  T_k &\simeq T^0_k + J_k \cdot V_k
  \quad,
\end{align}
where
\begin{align}
  T^0_k &= R^\dagger_{14}( 2 \eta_{0(k)} ) = R^\dagger_{14}( 2 \delta \Delta x_k )
  \nonumber{}\\
  &=
  \begin{pmatrix}
    \cos (2 \delta \Delta x_k) &0 &0 &-\sin (2 \delta \Delta x_k)
    \\
    0 &1 &0 &0
    \\
    0 &0 &1 &0
    \\
    \sin (2 \delta \Delta x_k) &0 &0 &\cos (2 \delta \Delta x_k)
  \end{pmatrix}
  \quad,
\end{align}
and
\begin{align}
  &J_k =
  \nonumber{}\\
  &\footnotesize{}
  \begin{pmatrix}
    2 \cos2\xi \sin(2 \delta \Delta x) \Delta x & \frac{ \sin2\xi \cos\varphi \sin(2 \delta \Delta x) }{\delta}
    & - \frac{ \sin2\xi \sin\varphi \sin(2 \delta \Delta x) }{\delta} & 2 \cos2\xi \cos(2 \delta \Delta x) \Delta x
    \\
    - \frac{ \sin2\xi \cos\varphi \sin(2 \delta \Delta x) }{\delta} &0 &0 & \frac{ 2\sin2\xi \cos\varphi \sin^2(\delta \Delta x) }{\delta}
    \\
    \frac{\sin2\xi \sin\varphi \sin(2 \delta \Delta x) }{\delta} &0 &0 & - \frac{ 2\sin2\xi \sin\varphi \sin^2(\delta \Delta x) }{\delta}
    \\
    -2 \cos2\xi \cos(2 \delta \Delta x) \Delta x & \frac{2\sin2\xi \cos\varphi \sin^2(\delta \Delta x) }{\delta}
    &- \frac{2\sin2\xi \sin\varphi \sin^2(\delta \Delta x) }{\delta} & 2 \cos2\xi \sin(2 \delta \Delta x) \Delta x
  \end{pmatrix}_{(k)}
  \,.
\end{align}
Then we have
\begin{align}
  \prod_{k=1}^n T_k &\simeq \prod_{k=1}^n ( T^0_k + J_k \cdot V_k )
  \nonumber{}\\
  &\simeq \prod_{k=1}^n T^0_k + \sum_{k=1}^n \left[ ( \prod_{i=1}^{k-1} T^0_i ) J_k ( \prod_{j=k+1}^n T^0_j ) \right] V_k
  \nonumber{}\\
  &= R^\dagger_{14}(\delta L_n) + \sum_{k=1}^n \left[ R^\dagger_{14}(\delta L_{k-1}) J_k R^\dagger_{14}\left( \delta(L_n - L_k) \right) \right] V_k
  \quad,
\end{align}
with $L_k$ defined in Eq.~(\ref{eq:x_k}).
According to Eq.~(\ref{eq:SVslab_T}), we only need to calculate the
first row of the $4\times4$ matrix,
whose elements are
\begin{align}
  Q_0 &= \cos\delta L_n + 2 \sin\delta L_n \sum_{k=1}^n \cos 2\xi_k  \cdot \Delta x_k V_k
  \nonumber{}\\
  Q_1 &= 2 \sum_{k=1}^n \sin2\xi_k \cos\varphi_k \cos\delta(L_{k-1} + \Delta x_k) \sin(\delta \Delta x_k) \cdot \frac{V_k}{\delta}
  \nonumber{}\\
  Q_2 &= -2 \sum_{k=1}^n \sin2\xi_k \sin\varphi_k \cos\delta(L_{k-1} + \Delta x_k) \sin(\delta \Delta x_k) \cdot \frac{V_k}{\delta}
  \nonumber{}\\
  Q_3 &= -\sin\delta L_n + 2 \cos\delta L_n \sum_{k=1}^n \cos2\xi_k \cdot \Delta x_k V_k
  \quad.
\end{align}
Finally,
from Eq.~(\ref{eq:ptb_Q}), we obtain the
approximation in Eq.~(\ref{eq:Ptb_slab_pert}) of $\Ptb(x_n)$ up to the
first order of $V$.

\subsection{Three-Neutrino mixing}
\label{sec:3-flavor}

In this subsection
we derive the limits of the expressions for
$\Ptot{E}{2}{e}$
that we have obtained in Eq.~(\ref{eq:Ptb_pert_V})
in the perturbative approximation
and
in Eq.~(\ref{eq:Ptb_slab_pert})
in the perturbative slab approximation
for a check of consistency with those derived,
respectively,
in Refs.~\cite{liao} and \cite{Valle}.

For $N_s=0$ and $\beta=e$, within standard parameterization
\cite{PDG} of the mixing matrix, we have
\begin{align}
      \theta_e = \theta_{12}\,,\quad\quad
      \chi_e=\theta_{13}\,,\quad\quad\phi_{e1} = \phi_{e2}=\varphi =0\,,
\end{align}
and therefore
\begin{align}
\xi =\theta_{12}\,,\quad\quad V = \half\cos^2\theta_{13}\Vcc\,,
\quad\quad \Ptb[V] = \cos^2\theta_{13}\sin^2\theta_{12}\,.
\end{align}

For the perturbative approximation, we have
\begin{align}
  \Ptot{E}{2}{e}(x_f) &= \Ptot{V}{2}{e} + \frac{ \cos^4\theta_{13} \sin^22\theta_{12} }{2}
  \intx \Vcc(x) \sin \left( 2 \Delta(x_f,x) \right) \dx
  \quad,\label{eq:P2e_pert_3f}
\end{align}
where $\Delta(x_f,x)$ is given by Eq.~(\ref{eq:Delta}) with
\begin{align}
  \dM &= \sqrt{ \left( \half \cos^2\theta_{13} \Vcc(x) - \delta \cos2\theta_{12} \right)^2 + ( \delta \sin2\theta_{12} )^2 }
  \quad.
\end{align}

For the slab approximation up to the first order of $\Vcc$, we have
\begin{align}
  \Ptot{E}{2}{e}(x_n) = \Ptot{V}{2}{e} + \frac{\cos^4\theta_{13} \sin^22\theta_{12} \sin( \delta L_n )}{2\delta}
  \sum^n_{k=1} \left( \sin \delta L_k - \sin \delta L_{k-1} \right) {\Vcc}_{(k)}\,.
  \label{eq:P2e_slab_3f}
\end{align}
The expressions in Eqs.~(\ref{eq:P2e_pert_3f}) and
(\ref{eq:P2e_slab_3f}) are consistent, respectively, with those
in Refs.~\cite{liao} and \cite{Valle}.

\section{Parametrization of $U$}
\label{sec:paramU}

The neutrino mixing matrix $U$ for the 3+1 mixing scheme can be
parametrized as the extension of the parametrization \cite{PDG} for
three-neutrino mixing
\begin{align}
  U=W^{34} W^{24} R^{14} R^{23} W^{13} R^{12}
  \quad,\label{eq:Uparam}
\end{align}
with $W^{ab}=W(\theta_{ab},\eta_{ab})$ being the complex unitary
matrix in the $(a,b)$ plane defined by
\begin{align}
[W(\theta_{ab},\eta_{ab})]_{rs}=\delta_{rs} &+
(\cos\theta_{ab}-1)(\delta_{ra}\delta_{sa} +
\delta_{rb}\delta_{sb})\nonumber{}\\
&+\sin\theta_{ab}(e^{-i\eta_{ab}}\delta_{ra}\delta_{sb}-e^{i\eta_{ab}}\delta_{rb}\delta_{sa})\quad,
\label{eq:Wab}
\end{align}
where $\theta_{ab}$ and $\eta_{ab}$ are the mixing angles and Dirac
CP-violating phases. The real unitary matrix $R^{ab}$ can be
obtained by setting $\eta_{ab}$ to be zero in $W^{ab}$.

In this parametrization, we can write down the explicit expressions
for the electron and sterile rows of the mixing matrix $U$:
\begin{align}
  \null & \null
  U_{e1} = c_{12} c_{13} c_{14}
  \,,
  \quad
  \null && \null
  U_{e2} = s_{12} c_{13} c_{14}
  \,,
  \label{Ue12}
  \\
  \null & \null
  U_{e3} = s_{13}e^{-i \eta_{13}} c_{14}
  \,,
  \quad
  \null && \null
  U_{e4} = s_{14}
  \,,
  \label{Ue34}
\end{align}
\begin{align}
  U_{s1} =& - s_{14} c_{12} c_{13} c_{24} c_{34}
  + ( s_{12} c_{23} + s_{13}e^{i\eta_{13}} s_{23} c_{12} ) s_{24} e^{i\eta_{24}} c_{34}
  \nonumber{}\\
  &+ (-s_{12} s_{23} + s_{13}e^{i\eta_{13}} c_{12} c_{23} ) s_{34} e^{i\eta_{34}}
  \,, \label{Us1}\\
  U_{s2} =& - s_{12} s_{14} c_{13} c_{24} c_{34}
  + (-c_{12} c_{23} + s_{12} s_{13}e^{i\eta_{13}} s_{23} ) s_{24} e^{i\eta_{24}} c_{34}
  \nonumber{}\\
  &+ ( s_{23} c_{12} + s_{12} s_{13}e^{i\eta_{13}} c_{23} ) s_{34} e^{i\eta_{34}}
  \,, \label{Us2}\\
  U_{s3} =& -s_{13}e^{-i\eta_{13}} s_{14} c_{24} c_{34}
  - ( s_{34} e^{i\eta_{34}} c_{23} + s_{23} s_{24} e^{i\eta_{24}} c_{34} ) c_{13}
  \,, \label{Us3}\\
  U_{s4} =& c_{14} c_{24} c_{34}
  \,. \label{Us34}
\end{align}
Moreover, we have
\begin{equation}
U_{\mu4} = c_{14} s_{24} e^{-i\eta_{24}} \,, \qquad U_{\tau4} =
c_{14} c_{24} s_{34} e^{-i\eta_{34}} \,. \label{Um4}
\end{equation}
We choose the following typical values for the oscillation
parameters as
\begin{align}
\rm{M1:}\quad\left\{\begin{array}{lllll}
\Delta m^2_{12} \simeq 7.54 \times 10^{-5}\,\mathrm{eV}\\
\theta_{12} \simeq 33.6^\circ\\
\theta_{23} \simeq 39.1^\circ\\
\theta_{13}\simeq 9.0^\circ\\
\theta_{14}=\theta_{24}=\theta_{34}=10^\circ\,,
\end{array}\right.\quad
\label{M1}
\end{align}
and
\begin{align}
\rm{P1:}\quad\begin{array}{l}
\eta_{13}=35^\circ\,,\quad\eta_{24}=75^\circ\,,\quad\eta_{34}=115^\circ\quad,
\end{array}
\label{P1}
\end{align}
where the three mixing angles ($\theta_{12}$, $\theta_{13}$ and
$\theta_{23}$) and $\Delta m^2_{12}$, equivalent to the active
neutrino oscillation parameters for three-neutrino mixing, are taken
from the global analysis \cite{fogli}, the three mixing angles
between the active and sterile flavors ($\theta_{14}$, $\theta_{24}$
and $\theta_{34}$) are motivated by the anomalies in the SBL data
\cite{LSND,Mini,R,Ga1,Ga2} and the non-trivial phases are selected
to illustrate the effects of the CP-violating phases.

\end{document}